# Big Data Approaches to Bovine Bioacoustics: A FAIR-Compliant Dataset and Scalable ML Framework for Precision Livestock Welfare

Mayuri Kate[1] and Suresh Neethirajan[1,2,*]
[1] Faculty of Computer Science, Dalhousie University, 6050 University Ave, Halifax, NS B3H 1W5, Canada
[2] Faculty of Agriculture, Dalhousie University, 62 Cumming Dr, Bible Hill, NS B2N 5E3, Canada

Correspondence*:
Suresh Neethirajan
sneethir@gmail.com

**Abstract**

The convergence of IoT sensing, edge computing, and machine learning is revolutionizing precision livestock farming. Yet bioacoustic data streams remain underexploited due to computational-complexity and ecological-validity challenges. We present one of the most comprehensive bovine vocalization datasets to date-569 expertly curated clips spanning 48 behavioral classes, recorded across three commercial dairy farms using multi-microphone arrays and expanded to 2,900 samples through domain-informed data augmentation. This FAIR-compliant resource addresses key Big Data challenges: volume (90 hours of raw recordings, 65.6 GB), variety (multi-farm, multi-zone acoustic environments), velocity (real-time processing requirements), and veracity (noise-robust feature-extraction pipelines). Our distributed processing architecture integrates advanced denoising (iZotope RX), multi-modal synchronization (audio-video alignment), and standardized feature engineering (24 acoustic descriptors via Praat, librosa, and openSMILE) to enable scalable welfare monitoring. Preliminary machine-learning benchmarks reveal distinct class-wise acoustic signatures across estrus detection, distress classification, and maternal-communication recognition. The dataset's ecological realism-embracing authentic barn acoustics rather than controlled conditions-ensures deployment-ready model development. This work establishes the foundation for animal-centered AI, where bioacoustic streams enable continuous, non-invasive welfare assessment at industrial scale. By releasing standardized pipelines with comprehensive metadata schemas, we advance reproducible research at the intersection of Big Data analytics, sustainable agriculture, and precision livestock management. The framework directly supports UN SDG 9, demonstrating how data science can transform traditional farming into intelligent, welfare-optimized production systems capable of meeting global food demands while maintaining ethical animal-care standards.

## 1 Introduction

The exponential growth of agricultural data—projected to surpass 5.1 exabytes by 2025—positions precision livestock farming at the intersection of IoT sensing, edge computing, and machine learning analytics [(Scoop Market Research 2025), (Precision Business Insights 2024)].

Within this evolving digital ecosystem, bioacoustic data streams stand out as a particularly complex and information-rich modality. These continuous, high-frequency temporal signals demand specialized preprocessing pipelines, robust feature engineering, and scalable analysis frameworks to unlock actionable insights. Building on this context, the rapid growth of digital agriculture has further highlighted the transformative potential of big data and machine learning in reshaping livestock farming into a more sustainable, welfare-centered, and efficient sector. Among various sensing modalities, bioacoustics has emerged as a powerful yet underutilized channel of information, offering non-invasive insights into animal health, behaviour, and emotional state. In particular, cattle vocalizations carry rich indicators of social interaction, estrus, maternal care, hunger, stress, and pain, positioning them as promising biomarkers for welfare monitoring and automated farm management systems. Harnessing these signals, however, requires curated datasets that faithfully capture the acoustic, behavioural, and environmental realities of commercial farming contexts.

The lack of large, annotated datasets remains one of the most significant bottlenecks in bovine bioacoustics research [(Kate and Neethirajan 2025)]. Traditional acoustic analysis methods involving manual spectrogram generation and feature extraction are informative but not scalable to the data volumes required for robust AI training, underscoring the critical need for comprehensive, FAIR (Findable, Accessible, Interoperable, Reusable)-compliant datasets that capture ecological validity while supporting big data analytics.

Despite increasing interest, existing bovine vocalization corpora remain limited in scale, scope, and reproducibility. Most prior datasets have been collected from small cohorts under controlled or homogeneous conditions, focusing primarily on a narrow set of call types such as estrus calls or distress vocalizations. For instance, the "BovineTalk" dataset reported over a thousand vocalizations but from only 20 cows in isolation, thereby excluding environmental noise and behavioural diversity. Similarly, physiological studies linking calls to cortisol or estrus relied on restricted conditions, limiting generalizability to commercial barns. These constraints hinder the development of machine learning models that can generalize across diverse farm environments, rare behaviours, and variable acoustic conditions. Furthermore, multimodal integration of audio with video or ethological annotations is rarely implemented, restricting opportunities to contextualize vocalizations with corresponding behaviours.

In addition to limited behavioural coverage, many existing datasets deliberately exclude background noise to ensure clean acoustic signals. While this simplifies analysis, it reduces ecological validity, as commercial dairy barns are acoustically complex environments containing mechanical noise, overlapping calls, and human activity. Models trained on clean laboratory recordings often fail when deployed in real-world farms, where vocal signals are embedded within heterogeneous soundscapes. There is therefore an urgent need for datasets that reflect the acoustic reality of farming environments, balancing signal clarity with ecological authenticity.

To address these gaps, we introduce a novel bovine vocalization dataset that combines scale, behavioural diversity, and ecological realism with rigorous annotation and metadata standards. The corpus comprises 569 curated clips spanning 48 behavioural classes, recorded across three commercial dairy farms in Atlantic Canada using a multi-microphone, multimodal design. By capturing audio simultaneously from multiple barn zones—feeding alleys, drinking troughs, milking parlours, and resting pens—and pairing these recordings with video observations and detailed ethological notes, the dataset provides a comprehensive representation of the acoustic and behavioural ecology of dairy cattle. Unlike earlier collections that prioritised controlled conditions, this resource embraces the complexity of barn environments, including background machinery, overlapping calls, and routine human activity, thereby enhancing its value for developing robust, field-ready analytical models.

A key contribution of this dataset lies in its ethology-driven annotation scheme, which organises vocalizations into nine main categories and 48 sub-types covering maternal, social, reproductive, feeding, drinking, handling, distress, environmental, and non-vocal events. Each clip is annotated with behavioural context, emotional valence, and confidence scores, enabling analyses that extend beyond acoustics to questions of welfare, motivation, and social interaction. This structure aligns with contemporary animal welfare frameworks that emphasise emotional valence and arousal, while also providing machine-readable descriptors suitable for computational modelling. Equally important is the dataset's adherence to FAIR principles. Metadata tables document recording context, equipment, clip features, and preprocessing parameters in a transparent and reproducible manner. The inclusion of acoustic features extracted with standardised pipelines (Praat, librosa, openSMILE) ensures interoperability with other livestock bioacoustic resources and facilitates downstream applications ranging from supervised classification to exploratory behavioural analysis.

Together, these elements establish this corpus as the most comprehensive and ecologically valid dataset of bovine vocalizations to date. It provides not only a foundation for advancing machine learning approaches to livestock sound analysis, but also a benchmark resource for researchers in animal behaviour, welfare science, and precision livestock management.

In addition to its methodological and scientific contributions, this dataset holds direct significance for the emerging field of precision livestock farming. By enabling the detection and interpretation of vocal cues linked to health, reproduction, and welfare, it opens pathways for non-invasive monitoring systems that can assist farmers in real time. Early detection of estrus, distress, or discomfort through automated vocal analysis could enhance reproductive management, reduce disease risks, and improve overall herd wellbeing. Beyond cattle, the dataset also contributes to the broader movement in animal-centered AI, where bioacoustic data are increasingly leveraged to give "digital voices" to non-human species.

The remainder of this paper is structured as follows. Section 2 presents the novelty of the dataset, situating it in relation to previous studies. Section 3 describes the data collection protocols, including recording sites, equipment, and multimodal capture methods. Section 4 outlines the preprocessing pipeline, covering noise profiling, filtering, denoising, segmentation, and annotation. Section 5 details dataset creation, including feature extraction, biological interpretation, metadata design, and preliminary analyses. Finally, Section 6 presents the discussion of significance and limitations, and Section 7 concludes the paper highlighting its potential as a benchmark for both animal welfare science and big data applications in agriculture. Building on these motivations, the next section outlines the novelty of our dataset in relation to existing bovine vocalization corpora.

## 2 Dataset Novelty

### 2.1 Scale and diversity of recordings

This work presents one of the most comprehensive bovine vocalization datasets to date. The corpus comprises 569 clips covering 48 behavioural labels (classes) and has a mean clip duration of ~ 21 s (median ~ 13.8 s; range 2.8–445 s). Analysis shows a long tailed distribution: the largest classes, Estrus_Call (117 clips) and Feed_Anticipation_Call (113 clips), account for 40% of the data, whereas many categories contain fewer than ten samples, reflecting the rare and spontaneous nature of some behaviours. Clip durations are short enough to facilitate fine grained acoustic analysis yet long enough to capture the full vocalization plus context.

Most published bovine call datasets are smaller both in scale and scope. For example, the "BovineTalk" study isolated 20 multiparous cows for 240 min post milking and obtained 1,144 vocalizations (952 high frequency and 192 low frequency) (Gavojdian et al. 2024); the authors noted that calls were recorded under identical conditions and excluded noise. Another study analysed 12 Holstein heifers and reported that vocalization rate peaked one interval before estrus climax and was higher during natural than induced estrus (Röttgen et al. 2018). (Yoshihara and Oya 2021) recorded 290 calls from 32 cows across four physiological states (feed anticipation, estrus, communication and parturition) and showed that call intensity, pitch and formant values reflected changes in salivary cortisol. (Katz 2020) captured 333 high frequency calls from 13 heifers and demonstrated that cows maintain individual vocal cues across contexts. Compared with these studies, our dataset contains both high and low frequency calls across positive and negative contexts, includes a richer set of behavioural classes (maternal, social, estrus, feeding, drinking, handling, distress, environmental and non vocal), and encompasses multiple farms and barn zones. This breadth enables analyses of behavioural diversity and cross context variation not previously possible.

Beyond scale and diversity, novelty also arises from the recording design, which is detailed in the next subsection.

### 2.2 Multi-Microphone Setup and Multimodal Synchronization

Recordings were collected from three commercial Holstein–Friesian dairy farms in Sussex County, New Brunswick, Canada over three consecutive days (5–7 May 2025), with one farm recorded per day between 9:00 am and 6:00 pm (Table 1). Across these sites, 65 raw audio files were obtained, representing a total of ~ 90 h of recordings (65.57 GB; mean file duration ~ 1 h 24 m, mean file size ~ 1.0 GB).

The dataset captures four primary behavioural contexts—drinking, feeding, milking, and resting—within commercial barn environments. These settings included natural background sounds such as clanging metal gates, fans, tractors idling, overlapping vocalizations, and other routine noises. By integrating multiple farms and barn zones, the corpus reflects the true acoustic diversity of commercial dairy environments rather than controlled or laboratory conditions.

The novelty of this corpus lies in its multimicrophone, multi-sensor design. Unlike earlier bovine vocalisation studies that relied on a single microphone or controlled settings, the present dataset integrates recordings from multiple farms, barn zones, and equipment types. Directional shotgun microphones (Sennheiser MKH 416, RØDE NTG-2) provided close-range, high signal-to-noise vocalisations, while portable recorders (Zoom H4n Pro, Zoom F6) and an autonomous Wildlife Bioacoustics logger captured longer-duration ambient soundscapes.

This design ensured that both focal vocal events (e.g., estrus calls, feeding anticipation) and background acoustic context (e.g., overlapping moos, barn machinery, human activity) were represented. The inclusion of video recordings from multiple GoPro cameras added a complementary visual dimension, enabling cross-referencing of acoustic events with behavioural context. Together, this multimodal, multi-sensor strategy produces a holistic, reproducible dataset that captures the acoustic reality of commercial dairy environments at unprecedented scale and resolution.

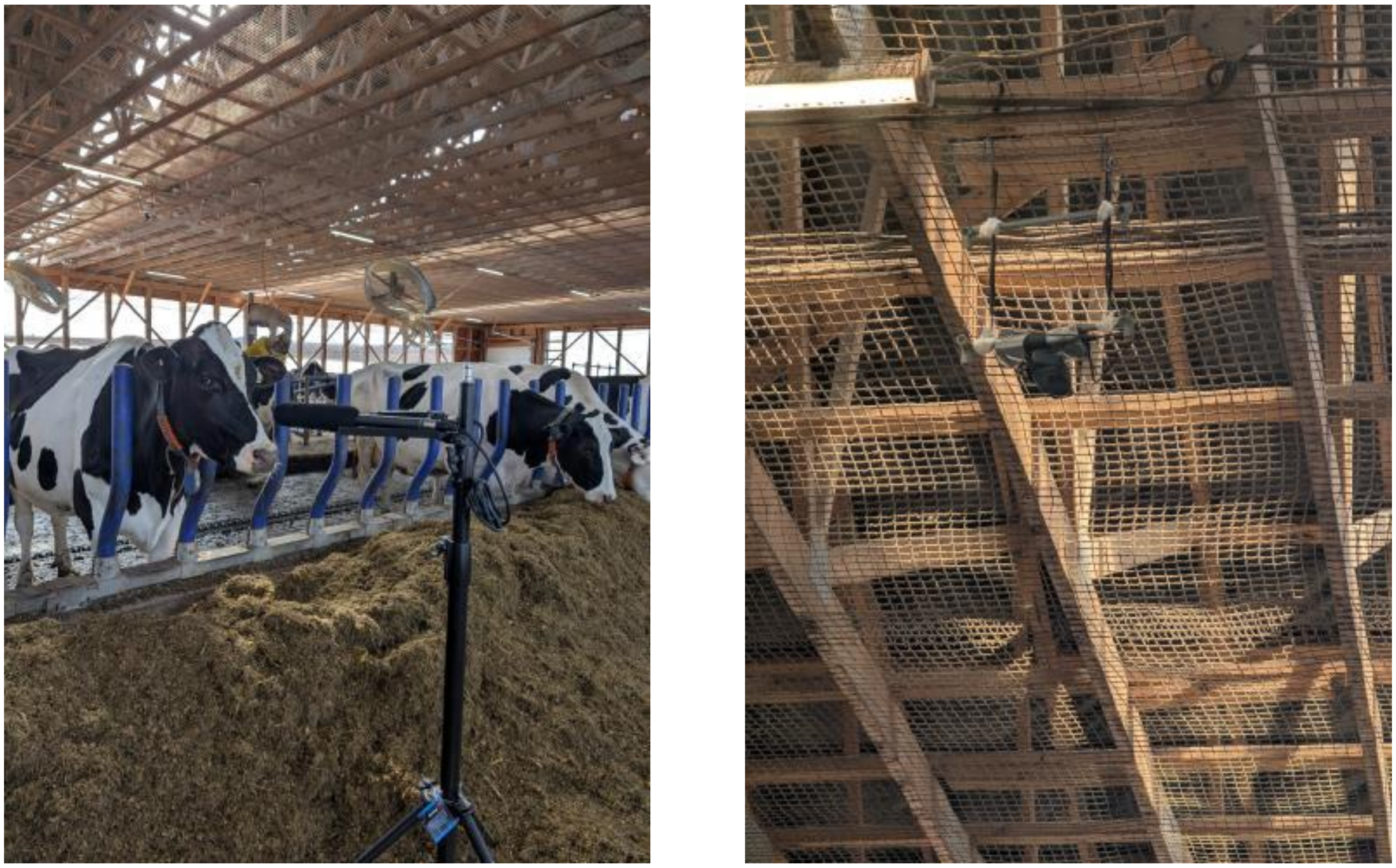

**(A)** Feeding Zone Microphone Setup **(B)** Feeding Zone Camera setup

***Figure 1. Multimicrophone and video setup used for synchronized multimodal recording in dairy barns.*** *(A) Directional RØDE NTG-2 shotgun microphone connected to a Zoom H4n Pro portable recorder, deployed in the feeding section of farm. The setup was mounted on a stable tripod oriented toward the feed trough to capture high-fidelity vocalizations while rejecting off-axis barn noise. (B) Ceiling-mounted GoPro action camera positioned directly above the same zone in same farm to capture continuous 4K video of feeding behaviour. The spatial alignment of microphone and camera ensured accurate cross-referencing between audio and behavioural context during manual annotation and ethological validation.*

### 2.3 Environmental noise profiling

Dairy barns are acoustically challenging, with persistent machinery noise (milking robots, feeders, tractors), metal gate clanging, hoof impacts, people talking, urination, and wind. To enable robust vocalisation extraction, separate noise recordings from each zone and analysed using the Welch method with a 16,384-sample FFT window. The resulting noise inventory (Table 3) summarises the spectral range and amplitude of different noise sources. Drinking noise exhibited low frequencies around 20 Hz and broad high-frequency peaks up to 1,029 Hz; the mean peak amplitude across samples was $\sim -40$ dB. Feeding noise (mix of hisses, horns, and gate impacts) had low frequencies from 30 Hz and high-frequency components up to ~567 Hz with similar amplitude. Milking noise from robotic equipment was dominated by low frequencies near 12 Hz and high frequencies up to ~300 Hz and was louder (mean amplitude $\sim -22$ dB). Resting noise (urination and human speech) spanned 12–493 Hz with mean amplitude $\sim -21$ dB. These profiles informed the design of a band-pass filter (150–800 Hz) to remove low-frequency machinery noise and high-frequency electrical hiss while preserving the vocalization band. By providing a quantitative noise inventory, our dataset allows researchers to reproduce the preprocessing pipeline and evaluate the robustness of acoustic features against background noise.

### 2.4 Annotation scheme and ethological foundation

A key novelty of the dataset is its detailed annotation scheme (Table 5) informed by ethological principles and welfare protocols. The annotation system organises vocalizations into nine main categories:

- Maternal & calf communication – includes calls such as Mother_Separation_Call (low frequency plaintive call when a cow is separated from her calf), Calf_Contact_Call (high frequency squeal when a calf seeks contact), and Maternal_Response_Call (mother cow responding back to her calves).

- Social recognition & interaction – encompasses affiliative calls like Greeting_Moo, Group_Contact_Call, Response_Exchange_Call, Herd_Coordination_Call and Social_Bonding, reflecting social hierarchy and cohesion.

- Estrus & mating behaviour – includes Estrus_Call, characterised by loud high frequency bellowing signalling sexual receptivity; Mating_Excitement_Call; and Mounting_Associated_Call.

- Feeding & hunger related – covers Feed_Anticipation_Call (calls before feeding), Feeding_Fustration_Call, and Chewing_Rumination_Sounds (non vocal chewing noises).

- Water & thirst related – includes Drinking_Slurping_Sounds, Water_Anticipation_Call and Hydration_Distress.

- Distress & pain – covers High_Frequency_Distress (intense distress vocalizations), Frustration_Call, Injury_Pain_Moo, Sneeze, Cough and Burp.

- Environmental & situational – includes vocal responses to environmental stimuli like Weather_Response_Call, Transportation_Stress_Call and Confinement_Protest_Call.

- Non vocal sounds – comprises non vocal behaviours such as Breathing_Respiratory_Sounds and Licking_Sounds.

Each sub category is accompanied by a concise description (e.g., Estrus_Call is a prolonged high frequency call emitted by a receptive female; Feed_Anticipation_Call is a rhythmic moo produced when cows expect feeding). The categories deliberately span positive and negative welfare states in line with welfare science. Prior research supports this ethological structuring: high frequency calls with the mouth open are associated with distress or long distance communication, whereas low frequency calls with the mouth closed occur in calm social contexts (Jobarteh et al. 2024). Vocalization rates also provide behavioural cues—for example, (Röttgen et al. 2018) found that call rate peaks before estrus climax , and (Yoshihara and Oya 2021) linked increased formant frequencies to elevated cortisol during parturition. By capturing a wide range of vocal types and non vocal sounds, our annotation scheme enables analysis of both emotional valence and arousal, as recommended in contemporary animal welfare frameworks.

### 2.5 Rich metadata and FAIR compliance

The dataset is accompanied by a comprehensive metadata table (Table 6) describing each clip. Key fields include the unique file name, recording date, farm identifier, barn zone, microphone model, duration, pitch statistics (mean, minimum and maximum), and formant frequencies (F1 and F2), along with categorical annotations such as main category, subcategory, emotional context, confidence score, and textual description. Feature definitions follow acoustic and ethological conventions—for example, pitch relates to laryngeal tension and arousal, formant spacing reflects vocal tract length, and energy measures indicate call strength.

The metadata structure aligns with the FAANG (Functional Annotation of Animal Genomes) guidelines for animal metadata and adheres to the FAIR principles (Harrison et al. 2018). These principles ensure that datasets are described with rich metadata, use community standards, and are stored in formats that facilitate long-term reuse across research communities. Applying FAIR to bioacoustic corpora enhances transparency, reproducibility, and integration with other animal genomics and welfare datasets (Harrison et al. 2018).

Overall, the dataset's novelty lies in its scale, diversity, multi-microphone design, multimodal referencing, detailed ethology-driven annotation scheme, and metadata structure that complies with international standards. Compared with previous bovine vocalization studies that focused on small cohorts or narrow behavioural contexts, this corpus offers a more holistic and reproducible resource for advancing machine learning and welfare research in dairy cattle. Having established the dataset's scope and comparative novelty, we now describe the data collection process, including recording sites, equipment, and behavioural context capture.

## 3 Data Collection

### 3.1 Recording sites

Data were collected from three commercial dairy farms in Sussex County, New Brunswick, coded FARM1–FARM3, all of which operated free-stall barns (Table 1). Farm 1 housed 57 Holstein cows, Farm 2 housed 207 cows in a mixed Holstein–Jersey herd, and Farm 3 housed 160 Holstein cows. Recording took place during sequential site visits (5–7 May 2025), ensuring coverage across farms under comparable seasonal conditions.

Each barn was divided into four monitored zones - drinking troughs, feeding alleys, milking parlour, and resting pens. Microphones were positioned at cow head height within each zone to optimise signal capture, for example above water bowls, near feed mangers, and along resting stalls. This arrangement allowed simultaneous, zone-specific recording, providing contrasts between different acoustic environments such as feeding areas, parlours, and resting pens.

In addition, manual observation logs were maintained throughout, noting events such as feeding schedules, veterinary visits, or machinery maintenance. These logs ensured that contextual events were linked to the acoustic data, creating a diverse soundscape that is representative of everyday husbandry practices in Atlantic Canadian dairy barns.

To capture these environments effectively, a multimicrophone hardware setup was deployed, as described below.

### 3.2 Recording hardware

To ensure representative acoustic coverage, a multimicrophone array was deployed, combining directional shotgun microphones with portable recorders and one autonomous bioacoustics logger:

- Sennheiser MKH 416 — hypercardioid interference-tube shotgun microphone, widely used in film and wildlife recording. Its strong side rejection allowed capture of subtle vocal nuances despite barn noise (Sennheiser, 2023).
  - Paired with Zoom F6 — six-channel portable recorder powered via phantom supply. The F6 supported 32-bit float recording, dual A/D converters, and ultra-low-noise preamps, preventing clipping even during high-intensity calls (Zoom, 2023).
- RØDE NTG-2 — supercardioid shotgun microphone, battery/phantom powered, valued for affordability and portability, suited for close-range recordings in drinking and milking contexts (RØDE, 2023).
  - Paired with Zoom H4n Pro — four-track handheld recorder powered by two internal AA rechargeable batteries. The H4n Pro included built-in X/Y stereo microphones, dual XLR inputs, and 24-bit recording with maximum SPL handling of 140 dB (Zoom, 2022). Two such RØDE NTG-2 + H4n Pro pairs were deployed for zone-specific coverage.
- Wildlife Bioacoustics autonomous recorder — a single passive logger programmed for scheduled monitoring, especially in resting areas. It operated continuously on four AAA rechargeable batteries, enabling long-duration capture without human presence (Wildlife Acoustics, 2023).

**Table 1. Overview of recording sites and microphone–recorder setup**

| Farm ID | Herd size (Holstein/others) | Barn zones monitored | Microphones | Recorders | Raw files (n) | Total duration (h) | Total size (GB) |
|---|---|---|---|---|---|---|---|
| **Farm 1** | 57 Holstein | Feeding, Drinking, | Sennheiser MKH 416; | Zoom F6; Zoom H4n | 21 | ~30 h | ~21.5 |

| Farm ID | Herd size (Holstein/others) | Barn zones monitored | Microphones | Recorders | Raw files (n) | Total duration (h) | Total size (GB) |
|---|---|---|---|---|---|---|---|
| | | Milking, Resting | RØDE NTG-2 | Pro | | | |
| **Farm 2** | 207 Holstein / Jersey mix | Feeding, Drinking, Resting, Milking | Sennheiser MKH 416; Zoom H4n Pro | Zoom F6; Zoom H4n Pro | 21 | ~29 h | ~22.0 |
| **Farm 3** | 160 Holstein | Feeding, Resting, Drinking | Wildlife Bioacoustics Recorder; RØDE NTG-2 | Zoom F6; Zoom H4n Pro | 23 | ~31 h | ~22.1 |
| **Total** | 424 cows | All zones (Feeding, Drinking, Milking, Resting) | Multiple (MKH 416, NTG-2, Bioacoustics) | Zoom F6; H4n Pro | 65 | 90 h 2 m | 65.6 |

**File characteristics:** Zoom F6 recordings were largest on average (~1.7 GB, mean peak amplitude −43.7 dB, 20–1,029 Hz range), followed by Zoom H4n Pro (~1.6 GB, −30.5 dB, 20–820 Hz). The autonomous logger produced smaller files (~308 MB, −20.1 dB, 12–893 Hz). By context, the drinking zone generated the largest raw data volume (~3.65 GB), followed by feeding (~1.7 GB), milking (~762 MB), and resting (~450 MB).

**Cameras:** Five fixed GoPro action cameras were installed above barn zones, recording continuously at 4K/30 fps with wide-angle lenses to cover feeding alleys, resting pens, drinking troughs, and the milking parlour. Cameras were synchronised with audio via a shared timecode feed.

**Placement strategy:** Microphones were mounted on adjustable stands ~ 1 m above cow head height, oriented toward the zone centre. This prevented contact with animals, minimised wall reflections, and ensured zone-specific capture. Cables were routed along beams and tripods stands and shielded to prevent chewing. The multimicrophone array (Fig. 2) enabled concurrent multi-zone recording and cross-comparison of calls across environments.

This combination of hardware provided complementary perspectives: phantom-powered Sennheiser + Zoom F6 setups for high-fidelity focal recording, AA-powered RØDE + Zoom H4n Pro pairs for flexible mobile coverage, and the AAA-powered Wildlife Acoustics logger for unattended long-term monitoring. Together with parallel video capture, the setup preserved both individual-level vocal features and group-level acoustic context, forming a robust foundation for behavioural and machine-learning analyses.

**Table 2. Microphone and recorder setup used for barn vocalization recordings.**

| Microphone / Recorder | Key specifications | Deployment zone(s) | Rationale in barn context | Avg. size (MB) | Avg. duration | Peak (dB) | Freq. range (Hz) |
|---|---|---|---|---|---|---|---|
| Sennheiser MKH 416 (shotgun) + Zoom F6 | Hypercardioid, interference tube; 40–20,000 Hz; paired with 32-bit float recorder | Feeding alleys, resting areas | Extremely directional; avoids barn noise and prevents clipping during loud moos | ~ 1,738 | ~ 1 h 28 m | –43.7 | 20–1,029 |
| RØDE NTG-2 (shotgun) + Zoom H4n Pro | Supercardioid; 20–20,000 Hz; paired with 24-bit recorder | Drinking troughs, milking parlour | Affordable, portable; good for close-up calls with reduced side interference | ~ 1,555 | ~ 1 h 25 m | –30.5 | 20–820 |
| Zoom H4n Pro (built-in XY + XLR inputs) | Handheld, stereo + external inputs; 24-bit/96 kHz | Feeding and milking | Mobile capture, flexible for focal recording | – | – | – | – |
| Zoom F6 (standalone channels) | 6-channel, 32-bit float, dual A/D converters | Feeding/resting | Long sessions with wide dynamic range | – | – | – | – |
| Wildlife Acoustics logger | Passive autonomous system; duty-cycled | Resting pens, background | Continuous scheduled monitoring without human presence | ~ 308 | ~ 1 h 15 m | –20.1 | 12–893 |

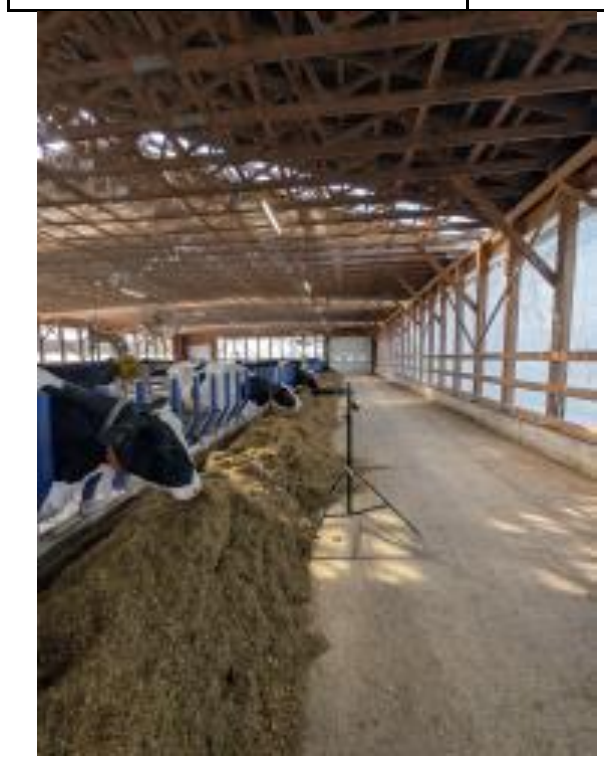

**(A)** RØDE NTG-2 near feeding alley

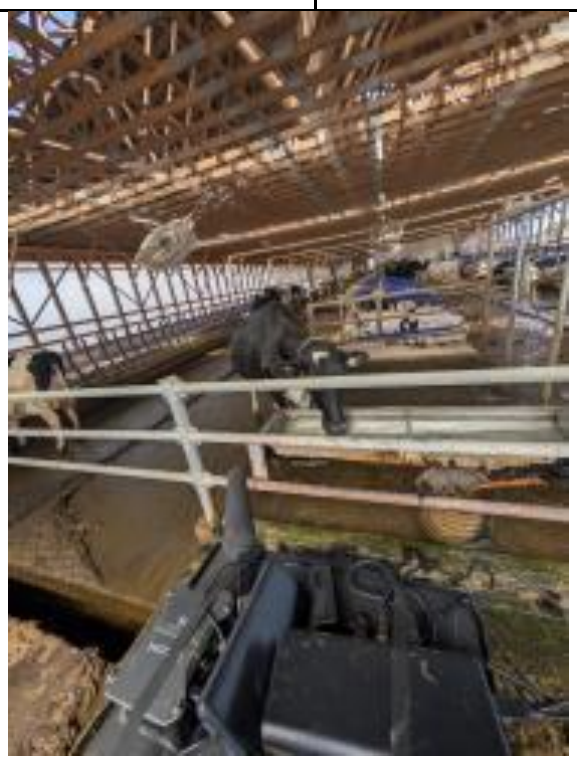

**(B)** Sennheiser MKH 416 near drinking trough

**(C)** Bioacoustics recorder in resting zone

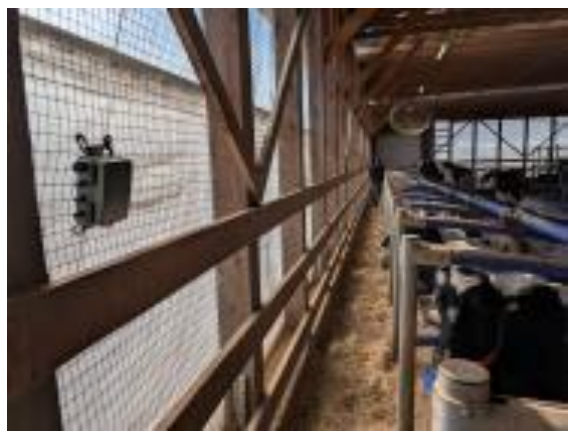

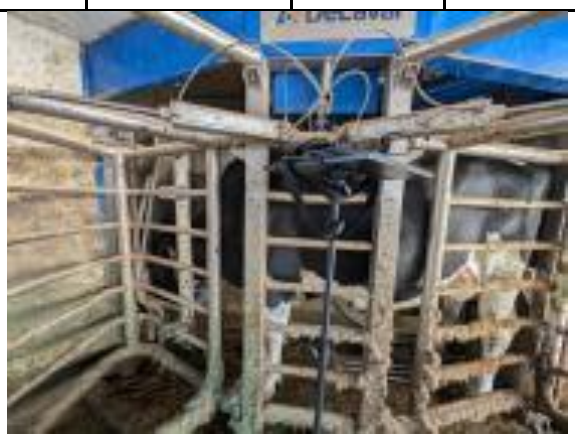

**(D)** RØDE NTG-2 with Zoom H4n Pro at milking station

***Figure 2. Microphone setups across barn zones at Farm 1.*** *(A) RØDE NTG-2 shotgun microphone positioned along the feeding alley to capture close-range vocalizations during feeding activity while minimizing side reflections. (B) Sennheiser MKH 416 directional*

*microphone placed near the drinking trough to record high-clarity vocal and non-vocal events amid tractors and metallic noise. (C) Autonomous Wildlife Bioacoustics recorder installed in the resting zone to capture low-frequency moos and background group vocalizations without human presence. (D) RØDE NTG-2 with Zoom H4n Pro handheld recorder positioned near the milking station to document vocal and non-vocal sounds associated with handling and milking routines.*

### 3.3 Behavioural context capture

Audio alone rarely conveys the full meaning of vocalisations. To provide behavioural context, the project implemented a multimodal capture protocol. Behavioural video was recorded using the fixed GoPro cameras described above, and three researchers maintained manual notes documenting the time of day, weather conditions, feeding schedules, milking events, and notable social interactions or stressors between cows. The annotation framework was informed by Tinbergen's four questions (Bateson and Laland 2013), which remain central in animal behaviour research. In the context of cow vocalisation, these can be adapted as follows:

- **Function** (adaptive value): What role does a call serve in daily life? For example, does it facilitate feeding coordination, signal distress, or attract social attention?
- **Phylogeny** (evolutionary background): How do vocal traits in dairy cattle relate to those seen in other bovids or domesticated species?
- **Mechanism** (causation): What immediate physiological or environmental factors trigger a vocalisation (e.g., hunger, pain, separation, handling, or barn noise)?
- **Ontogeny** (development): How does vocal behaviour vary with age, parity, or experience (e.g., heifers vs multiparous cows)?

By linking proximate mechanisms (physiology, environment) with ultimate functions (communication, adaptation), this framework strengthens interpretation of the dataset beyond acoustics alone. For instance, high-frequency open-mouth calls may be tied to immediate arousal or stress, while also serving long-term communicative roles within the herd (Jobarteh et al. 2024).

Audio–video alignment was performed manually. Instead of using automated synchronisation hardware, researchers cross-referenced the timestamps of audio recordings with video footage and their own observation logs. This practical approach enabled vocal events to be matched with visible behaviours (such as feeding, resting, or responding to handling) without specialised tools.

### 3.4 File handling and storage

Field data were initially stored on the internal memory cards of the Zoom recorders and GoPro cameras. To prevent overwriting or accidental data loss, recordings from each farm were transferred immediately after the day's data collection. Files were copied to a secure laptop on-site and then uploaded to a shared Dropbox repository, ensuring both immediate backup and remote accessibility. Audio recordings were saved in WAV format (44.1 or 48 kHz, 24-bit depth) and named according to a structured convention: FarmID – MicrophonePlacement – BarnZone – Date – Time. Video files were stored in MP4 format with matching time stamps to maintain cross-referencing with audio.

To safeguard data integrity, the raw original files remain archived in Dropbox. For subsequent analysis steps such as preprocessing, cleaning, and segmentation, working copies were downloaded and processed locally, ensuring that the original dataset was preserved without modification. Metadata spreadsheets were updated during each transfer to log file names, times, equipment used, and backup status. This multi-stage handling approach—memory card → laptop → cloud backup → working copies—provided a robust and traceable workflow that minimised the risk of data loss and maintained strict separation between raw and processed datasets.

### 3.5 Big Data Processing Architecture

To efficiently manage and analyse the ~ 90 h of multimodal recordings (65 raw files; 65.6 GB), a modular and reproducible big-data workflow was established in alignment with FAIR and FAANG principles. The architecture (Fig. 3) integrates five sequential stages, from ingestion to analytics, ensuring traceable and scalable processing of livestock bioacoustic data.

- **Data Ingestion** – Multi-sensor audio and video streams were organized using a structured, time-stamped file-naming convention encoding farm, barn zone, date, time, and device ID. All recordings were mirrored between local and cloud repositories (Dropbox) to ensure redundancy, traceability, and immediate accessibility. Comprehensive metadata logs documented instrument settings, environmental context, and acquisition events.

- **Preprocessing** – Raw recordings were processed in batches following the standardized steps outlined in Sections 4.1–4.4, including spectral noise profiling, band-pass filtering, adaptive denoising, manual segmentation, and acoustic verification. Each transformation was logged to maintain reproducibility and quality control, yielding denoised, analysis-ready clips that preserved the ecological realism of barn soundscapes.

- **Feature Engineering** – For every segmented clip, a 24-feature acoustic vector was derived using Praat/Parselmouth, librosa, and openSMILE. These temporal, spectral, and cepstral descriptors capture biologically interpretable properties of vocalizations and serve as standardized inputs for statistical and machine-learning analyses.

- **Storage and Metadata Indexing** – Cleaned audio files, feature tables, and contextual metadata were stored in a hierarchical cloud structure maintaining explicit lineage across raw, processed, and derived assets. Metadata tables recorded preprocessing parameters, feature definitions, and annotation details, supporting long-term reusability and transparent provenance.

- **Analytics and Output Generation** – Feature tables were aggregated to produce statistical summaries, exploratory visualizations, and model-ready matrices for behavioural and acoustic analyses. The modular architecture enables efficient reprocessing when parameters are updated and facilitates seamless integration of future datasets or sensing modalities.

With raw recordings secured across farms and barn zones, the following section 4 details the preprocessing pipeline applied to enhance signal quality and prepare clips for segmentation and annotation.

**Ethical Approvals**

All experimental procedures were reviewed and approved by the Dalhousie University Animal Ethics Committee (Protocol No. 2024–026). Data collection involved no physical interaction with animals, and all participating farm owners were fully informed of the study's objectives and provided written consent. In accordance with institutional and national ethical standards, data were obtained solely through passive audio, image, and video recordings.

# 4 Data Preprocessing

## 4.1 Noise profiling

Noise spectral profiling was carried out as the first stage of preprocessing, since raw barn recordings contained a wide range of background sounds from machinery, metal gates, hoof impacts, people, and other animals. This process involved analysing background noise patterns in terms of frequency (Hz) and amplitude (dB) in order to distinguish cow vocalisations from environmental sources. Noise-only segments were extracted from the recordings for each barn zone and analysed using Audacity with the Welch spectrum function, configured with a 16,384-point FFT window, 50% overlap, and a logarithmic frequency axis. The Welch method was chosen because it averages overlapping segments, giving smoother spectra and suppressing transient spikes. Alternative methods such as Bartlett (less smoothing), Blackman–Harris (suited for controlled studio audio), and Hanning (unstable under noisy barn conditions) were considered, but Welch proved to be the most reliable for real-world farm recordings. The analysis revealed distinct noise signatures for different barn zones.

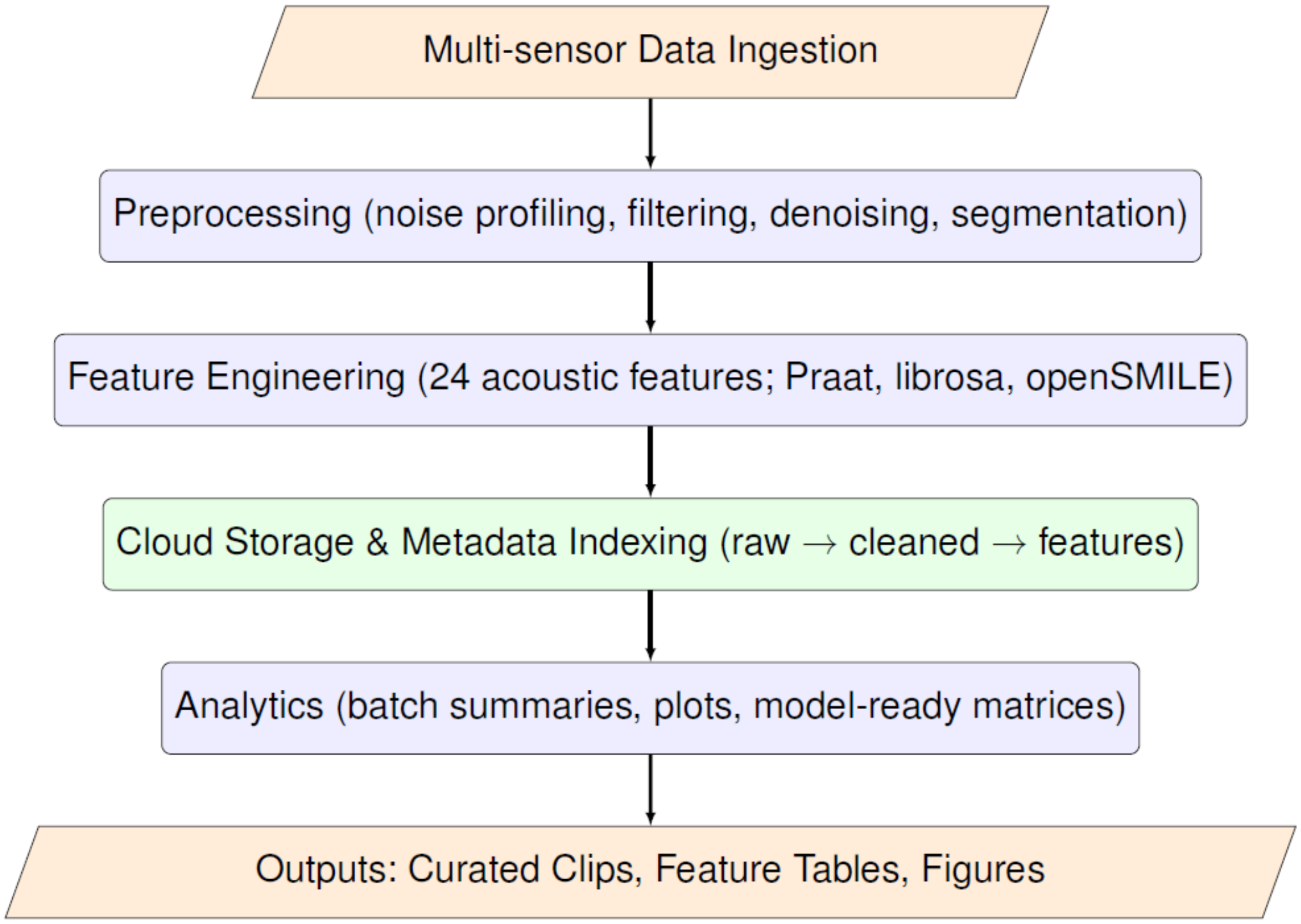

***Figure 3. Distributed big data processing architecture for multimodal livestock bioacoustics.*** *The pipeline integrates multi-sensor ingestion, parallel preprocessing, batch feature extraction, cloud storage, and scalable analytics supporting both batch and streaming computation*

- Drinking areas were dominated by metallic clanging of bowls, splashing at troughs, and bowel noise, with frequency peaks extending up to around 1029 Hz and mean amplitudes of approximately –60 dB.

- Feeding zones produced a mixture of metallic impacts, compressed air hisses, and horn-like sounds, spanning 30–300 Hz with variable amplitudes.

- Milking parlours were characterised by robotic systems, pumps, and vacuum lines, which generated relatively narrow frequency bands around 100–200 Hz but at higher amplitudes between –10 and –36 dB.

- Resting areas contained low-frequency components between 25–80 Hz from urination, rumination, and equipment hum, combined with higher-frequency sounds such as people talking.

- In addition, microphone hiss was consistently detected below 50 Hz across all farms and zones.

These findings were consistent with the expectation that barn-specific activities and equipment each contribute distinctive background noise signatures that overlap with the vocal frequency space of cows.

The profiles are summarised in Table 3, which presents the farm-wise and barn-zone-wise distribution of noise sources, frequency ranges, amplitudes, and the microphones used. This information directly informed the design of the filtering pipeline applied in subsequent preprocessing, where band-pass filtering was configured to retain the main vocal range (~ 100–1800 Hz) while attenuating machinery hum, human speech, and high-frequency hiss. A representative spectrogram and waveform comparison (Fig. 4) contrasts a cow vocalisation with a noise-only segment, illustrating the necessity of noise profiling before segmentation and feature extraction.

**Table 3. Farm-wise noise spectral profiles across barn zones.**

| **Day** | **Farm** | **Category** | **Noise Source** | **Low Frequency** | **High Frequency** | **Peak Amplitude** | **Microphone** |
|---|---|---|---|---|---|---|---|
| Day 1 | Farm 1 | Drinking | Metallic plates | 190 Hz | 655 Hz | −60.5 dB | Zoom F6 |
| Day 1 | Farm 1 | Feeding | Bowel movement | 229 Hz | 567 Hz | −54.8 dB | Zoom H4n |
| Day 1 | Farm 1 | Milking | Robotic milking system | 106 Hz | 127 Hz | −32.2 dB | Zoom H4n |
| Day 1 | Farm 1 | Resting | Tractor | 103 Hz | 120 Hz | −1.8 dB | Bioacoustics |

| Day | Farm | Category | Noise Source | Low Frequency | High Frequency | Peak Amplitude | Microphone |
|---|---|---|---|---|---|---|---|
| Day 2 | Farm 2 | Drinking | Birds chirping | 421 Hz | 579 Hz | −25.7 dB | Zoom H4n |
| Day 2 | Farm 2 | Feeding | Metal plates | 164 Hz | 283 Hz | −58.2 dB | Zoom F6 |
| Day 2 | Farm 2 | Milking | Robotic milking system | 217 Hz | 301 Hz | −25.7 dB | Zoom H4n |
| Day 2 | Farm 2 | Resting | Microphone hiss | 12 Hz | 19 Hz | −13.9 dB | Bioacoustics |
| Day 3 | Farm 3 | Drinking | Hiss + urination sound | 91 Hz | 146 Hz | −24.6 dB | Zoom H4n |
| Day 3 | Farm 3 | Feeding | Hiss + tractor horn | 30 Hz | 72 Hz | −14.2 dB | Zoom F6 |
| Day 3 | Farm 3 | Milking | People talk + milking system | 100 Hz | 130 Hz | −5.4 dB | Bioacoustics |
| Day 3 | Farm 3 | Resting | Hiss | 25 Hz | 38 Hz | −20.3 dB | Zoom H4n |

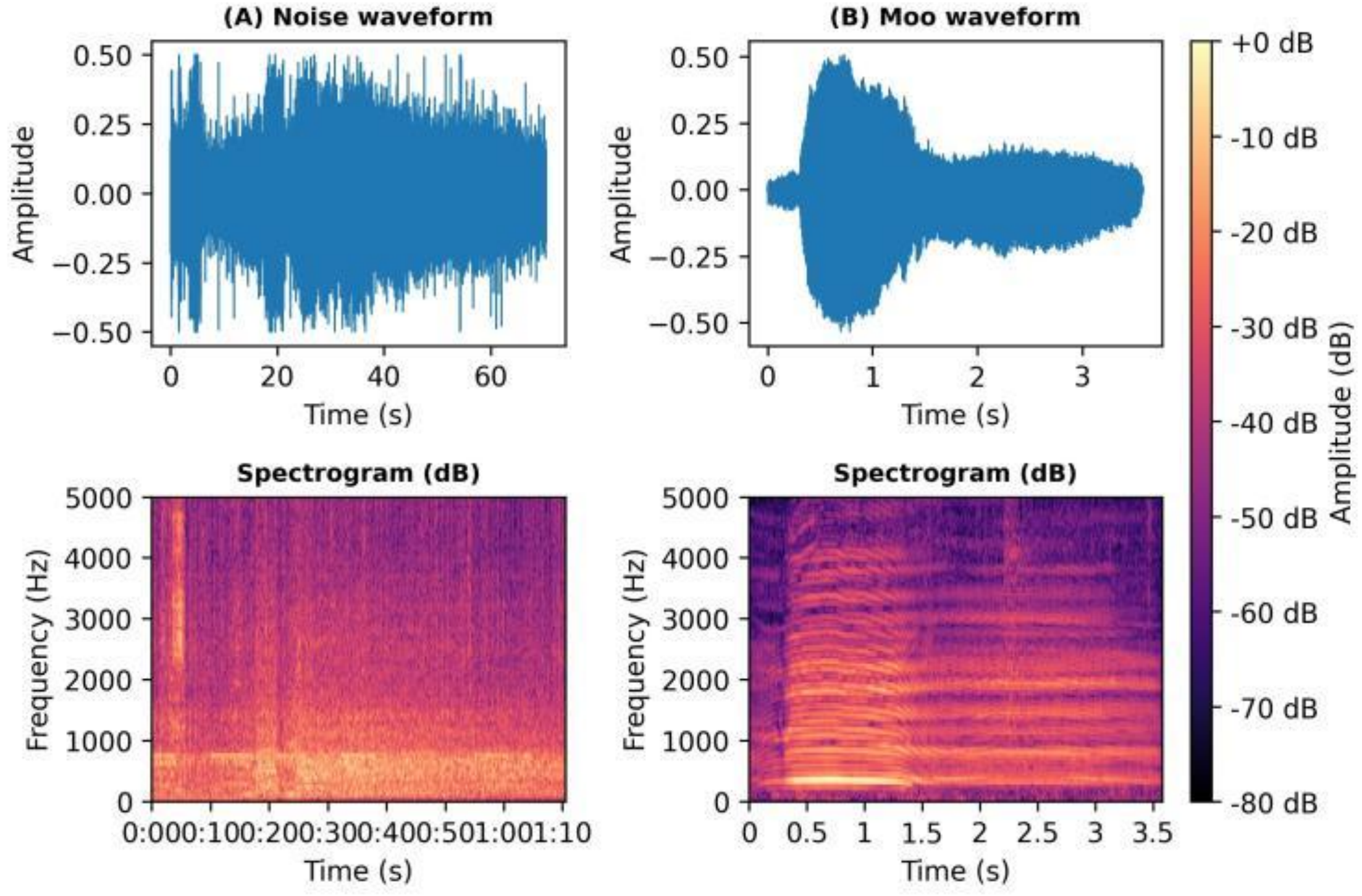

***Figure 4. Comparison of barn noise and cow vocalization spectrograms.*** *(A) Waveform and spectrogram of a barn noise segment dominated by broadband mechanical and environmental energy. (B) Waveform and spectrogram of a cow "moo" call recorded in the same acoustic environment, showing harmonic stacks and formant bands with periodic energy peaks.*

*Spectrograms were computed using short-time Fourier transform and displayed in decibel scale relative to the maximum amplitude to enable direct visual comparison of spectral structure and signal clarity.*

### 4.2 Band pass filtering

Based on the results of the noise spectral profiling and the known frequency ranges of bovine vocalizations, a fourth-order Butterworth band-pass filter was applied with cut-off frequencies set at 50 Hz and 1800 Hz. This frequency range was selected to retain the majority of bovine vocal energy—where the fundamental frequency typically lies between 100 and 300 Hz and harmonics extend up to approximately 1000 Hz—while attenuating low-frequency machinery hum below 50 Hz and high-frequency electrical hiss above 1800 Hz. Previous studies have reported comparable ranges, noting that cattle vocalizations commonly exhibit fundamental frequencies between 80 and 180 Hz for cows and calves, with energy extending up to 1 kHz or higher in certain contexts ((Torre et al. 2015), (Briefer 2012), (Lenner et al. 2025)). The Butterworth filter was chosen due to its maximally flat frequency response in the passband, which avoids distortion of harmonic structure and preserves acoustic fidelity. Its use is well established in acoustic and bioacoustic research as a robust method for isolating biologically relevant frequency bands ([(MacCallum et al. 2011), (Simmons et al. 2022)). To eliminate phase shifts that might affect later acoustic feature extraction, such as pitch contour or spectral energy analysis, filtering was implemented using zero-phase forward–backward filtering with the *butter* and *filtfilt* functions from Python's *scipy.signal* library.

This filtering step plays a critical role in the preprocessing pipeline. By suppressing spectral energy outside the vocalization band, it reduces broadband interference and makes the harmonic patterns of vocalizations stand out more clearly in spectrograms. This is particularly important in barn environments where background noise from ventilation systems, metallic clanging, and robotic milking machinery often overlaps with vocal frequencies in the 50–1800 Hz range. While some overlap remains, the band-pass filtering substantially improves the signal-to-noise ratio, emphasizing the stable formants of cow vocalizations, which are most prominent between 200 and 400 Hz ((Watts and Stookey 2000), (Green et al. 2019)).

The filtering process also generates two outputs. First, spectrogram data are exported as a CSV file, providing frequency bins over time in tabular form, which enables automated detection of vocal events such as moos, sneezes, or coughs without continuous manual listening. Second, filtered WAV audio files are produced, which present reduced background hiss and rumble and therefore facilitate cleaner listening and manual annotation. Together, these outputs provide both a human-audible and machine-readable foundation for subsequent stages of analysis.

Although this step does not entirely eliminate overlapping barn noise within the vocalization band, it substantially improves clarity and prepares the audio for further visualization, labeling, and feature extraction. A side-by-side comparison of pre- and post-filtering spectrograms (Fig. 5) illustrates the suppression of broadband noise while preserving the harmonic structure of cow calls.

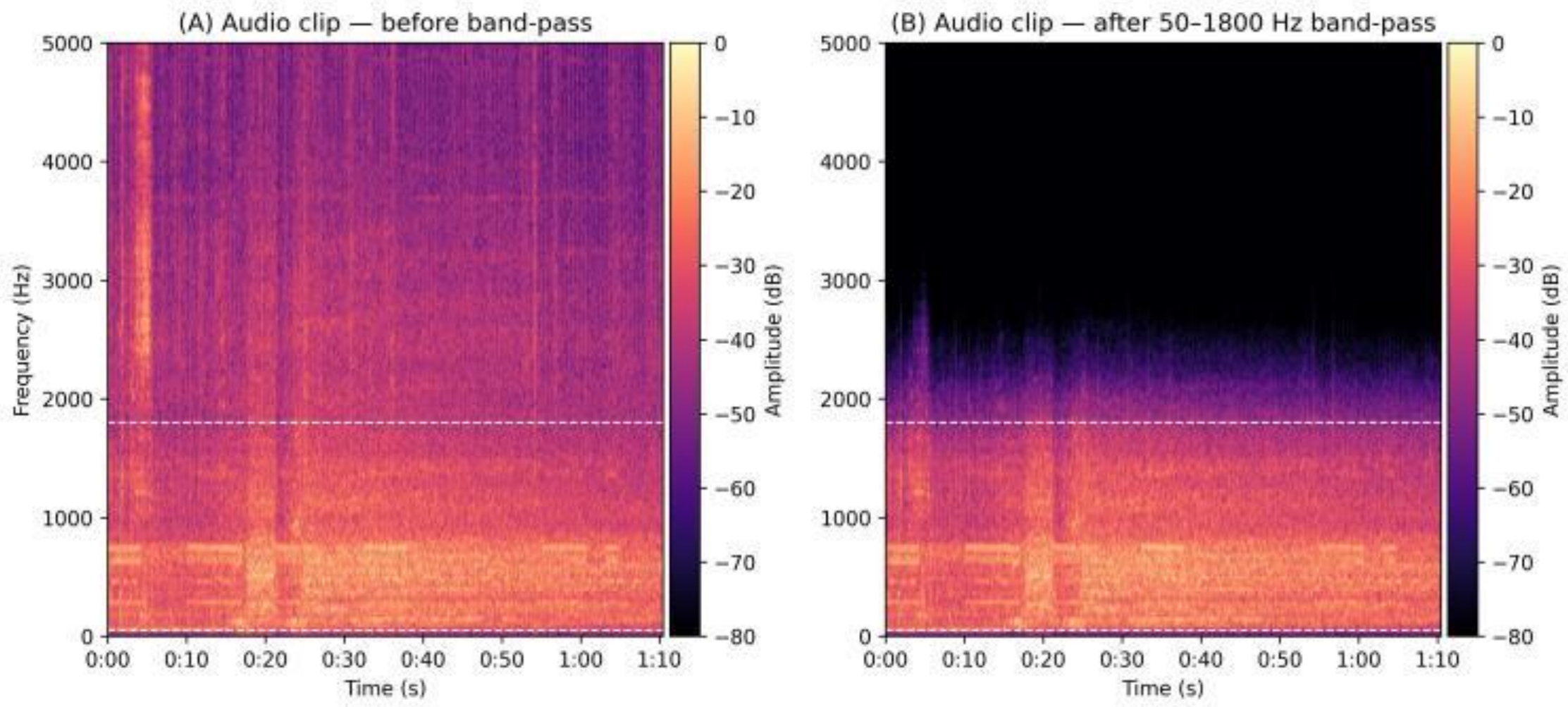

***Figure 5. Effect of band-pass filtering on an audio clip.*** *(A) Spectrogram of the unfiltered signal showing broadband energy, including low-frequency hum and high-frequency hiss. (B) Spectrogram after 50–1800 Hz band-pass filtering, with clear attenuation of energy below 50 Hz and above 1.8 kHz while preserving in-band harmonic structure. Dashed lines mark the passband; both panels share identical time and frequency axes (displayed to 5 kHz) and a common dB color scale for direct visual comparison.*

### 4.3 Noise reduction in iZotope RX 11

Following band-pass filtering, additional noise suppression was carried out using iZotope RX 11 (iZotope Inc., Cambridge, MA), a professional-grade audio restoration suite that offers fine-grained spectral editing, adaptive noise learning, and fault repair modules. Although iZotope RX is not yet cited in published bioacoustic studies, its functionality parallels classical spectral denoising and repair techniques long used in animal sound research (e.g. spectral subtraction, MMSE spectral estimators, wavelet denoising) ((Xie et al. 2021),(Brown et al. 2017), (Juodakis and Marsland 2022)). More recently, methods such as Biodenoising adopt high-quality pre-denoising (often via speech-based or spectral tools) as pseudo-clean references for training animal-specific models, which further validates the use of advanced denoising tools upstream. RX 11 was chosen because it allows the user to visually inspect the time–frequency structure, select noise-only regions, build adaptive noise profiles, and non-destructively subtract noise while preserving the vocal harmonics of interest. This flexibility is especially useful in barn settings, where noise is heterogeneous (e.g. ventilation hum, mechanical clanks, electrical hiss) and overlaps in frequency with vocal energy.

We implemented a multi-stage noise reduction pipeline consisting of the following steps:

1. Gain normalization and DC offset removal to standardize amplitude baselines across recordings.

2. Spectral De-noise to compute adaptive noise estimates from silent (non-vocal) segments and subtract them from the signal.

3. Spectral Repair to mitigate transient artifacts (e.g., gate slams or rapid mechanical clicks) by interpolating across missing time–frequency bins.

4. De-clip and De-crackle modules to correct occasional saturation and impulsive noise events.

5. EQ Match to smooth the overall frequency response and compensate for microphone coloration, resulting in a more natural tonal balance.

This pipeline produced a notable improvement in signal-to-noise ratio, such that cow vocalizations became more distinct in spectrograms, with cleaner harmonic continuity and fewer artifact interruptions. The enhanced clarity aids both manual annotation and downstream segmentation or feature extraction. A screenshot of the RX 11 interface showing a spectral editing session (Fig. 6) visually demonstrates how noise is isolated and removed while maintaining vocal structure. Although the approach does not fully eliminate overlapping noise within the vocal band, it significantly reduces interference and lays the foundation for robust event detection and feature analysis.

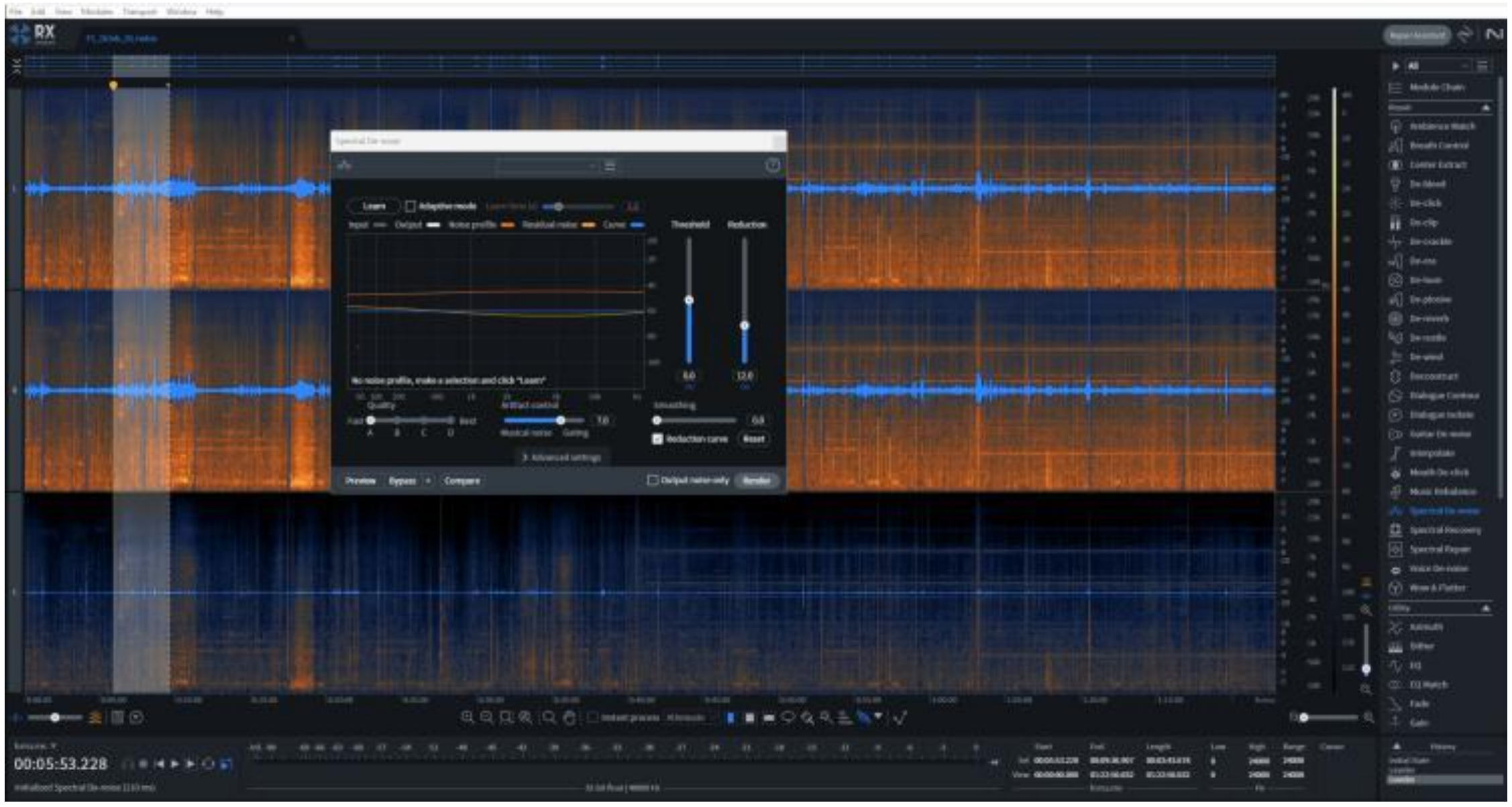

***Figure 6. Spectral denoising of a cow vocalization in iZotope RX.*** *Screenshot from iZotope RX (Spectral Denoise module) showing the noise-reduction workflow applied to raw barn audio prior to feature extraction. The upper panel displays the original spectrogram with broadband background noise typical of fan and machinery hum, while the lower panel shows the cleaned signal after adaptive spectral subtraction. Noise profile estimation was performed using a noise-only segment, and attenuation settings were tuned to preserve vocal harmonics while suppressing stationary low-frequency noise and high-frequency hiss. This preprocessing step enhanced signal-to-noise ratio and ensured clearer spectral features for subsequent analysis.*

### 4.4 Segmentation with Raven Lite and Acoustic Inspection in Praat

After denoising in iZotope RX 11, we segmented the continuous recordings into individual vocalizations using Raven Lite (Version 2.0). Raven Lite is a free audio analysis tool developed by the Cornell Lab of Ornithology and widely applied in ecological and behavioral bioacoustic studies (Dugan et al. 2016).Raven Lite was selected because it provides real-time waveform and spectrogram visualization with straightforward selection tools for manual clipping and export, and it is freely available and widely used in bioacoustic workflows. Although it lacks automated detection and batch-processing features available in Raven Pro, it was well-suited for the context-aware, manual segmentation required in this study.

We configured Raven Lite to display a short-time Fourier transform (STFT) spectrogram with a 1,024-point window, 50% overlap, and a Hamming window, (Fig. 7) which strikes a practical balance between time and frequency resolution for cattle calls. Candidate events were identified by visually scanning the spectrogram for harmonic stacks or broadband bursts and confirming each event by listening. Each clip was then extracted with start–end markers, and 2–3 s of padding were included on either side to preserve contextual cues (e.g., pre-onset inhalation, resonance tails). Onset was marked where amplitude rose above the noise floor and the first harmonic band became visible; offset was marked where energy returned to baseline—criteria consistent with common bioacoustic segmentation practice in livestock vocal studies. (Meen et al. 2015)

To maintain traceability, each selection was exported as a new WAV file and named with a structured convention encoding farm, zone, date, time, microphone , and a provisional class placeholder. Approximately 569 clips were extracted from  90 h of recordings. To reduce subjectivity, an independent second annotator cross-checked a subset of clips for boundary placement and completeness (inter-observer calibration), and we adopted a conservative policy of retaining extra seconds of context when uncertain.

Following clipping, we performed acoustic inspection in Praat (Boersma 2011) to verify segmentation boundaries and confirm that selections represented bona fide vocalizations rather than residual noise or mechanical transients (Green et al. 2019). In Praat (Fig. 8), we inspected pitch contours (F0) and formant tracks (F1, F2) alongside intensity envelopes to check that harmonic structure was continuous within marked boundaries and aligned with cattle-vocal production expectations reported in prior work. We then bridged Praat to Python via Parselmouth (Jadoul et al. 2018) for scripted extraction of Praat-native measures (e.g., pitch range, formants, harmonic-to-noise ratio), and complemented these with librosa features (e.g., spectral centroid, bandwidth, roll-off, zero-crossing rate) for machine-readable descriptors used later in modeling. This combination ensured consistency between human-verified boundaries and algorithmic features.

Raven Lite's strengths at this stage are its stability, clear spectrogram interface, and low operational overhead for manual, context-aware segmentation; limitations include the lack of batch annotation and advanced detectors, which exist in Raven Pro and in research code. As the dataset expands, we anticipate scaling via semi-automated detectors (e.g., HMM/ML pipelines or Raven Pro's detection tooling) as demonstrated in related cattle-vocal detection studies. For the present work, however, manual segmentation with conservative padding and cross-validation produced a high-confidence corpus appropriate for downstream feature extraction and analysis.

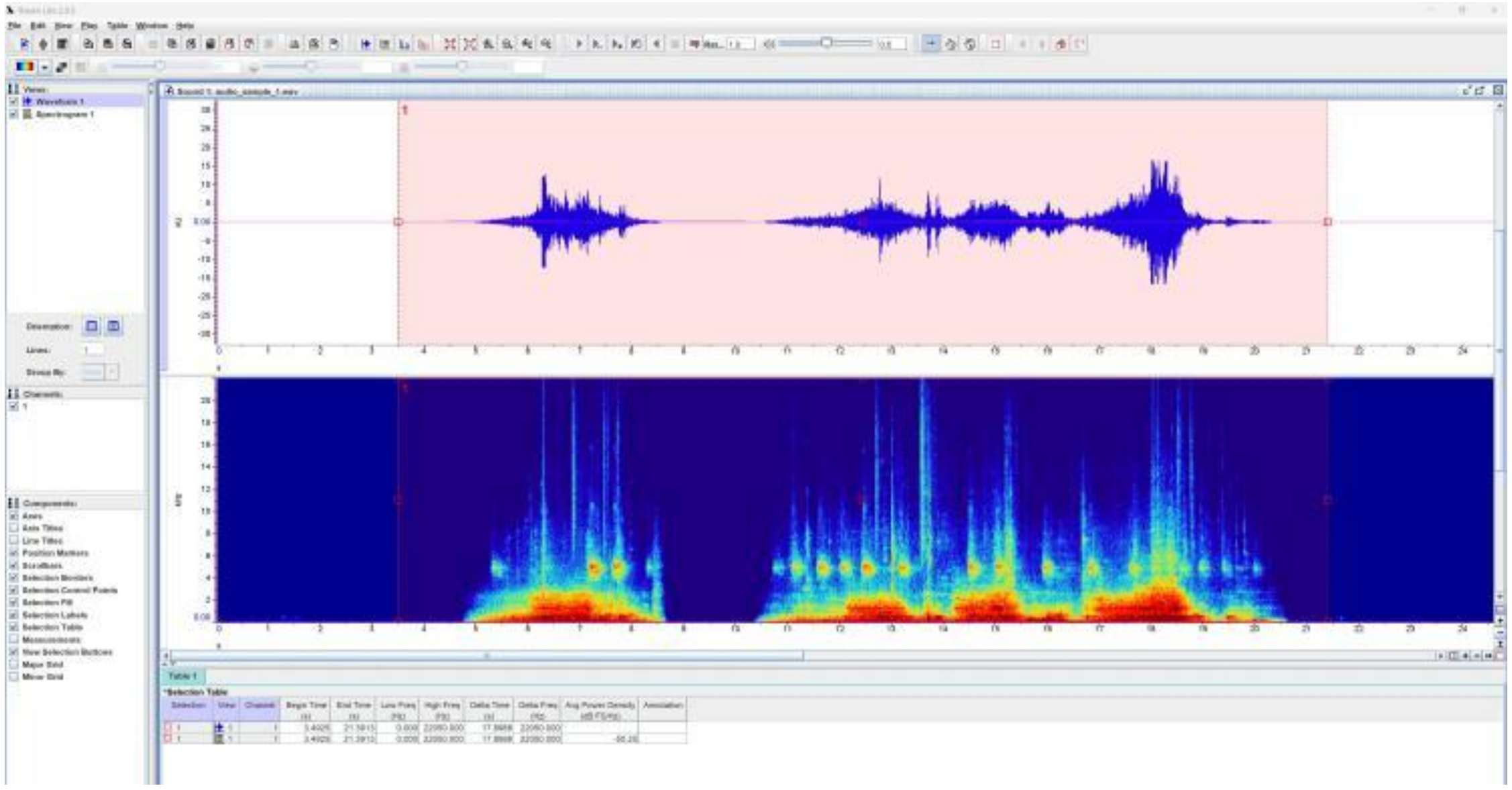

***Figure 7. Annotation of cow vocalizations using Raven Lite.*** *Screenshot from Raven Lite showing manual annotation of vocalization segments on the spectrogram view. Each selection box corresponds to an identified call event, with start and end times marked based on visible harmonic onset and offset boundaries. This visual verification ensured accurate segmentation of vocalizations and exclusion of mechanical or environmental noise. Annotated time–frequency regions were later used as reference intervals for automated feature extraction and labeling in Python.*

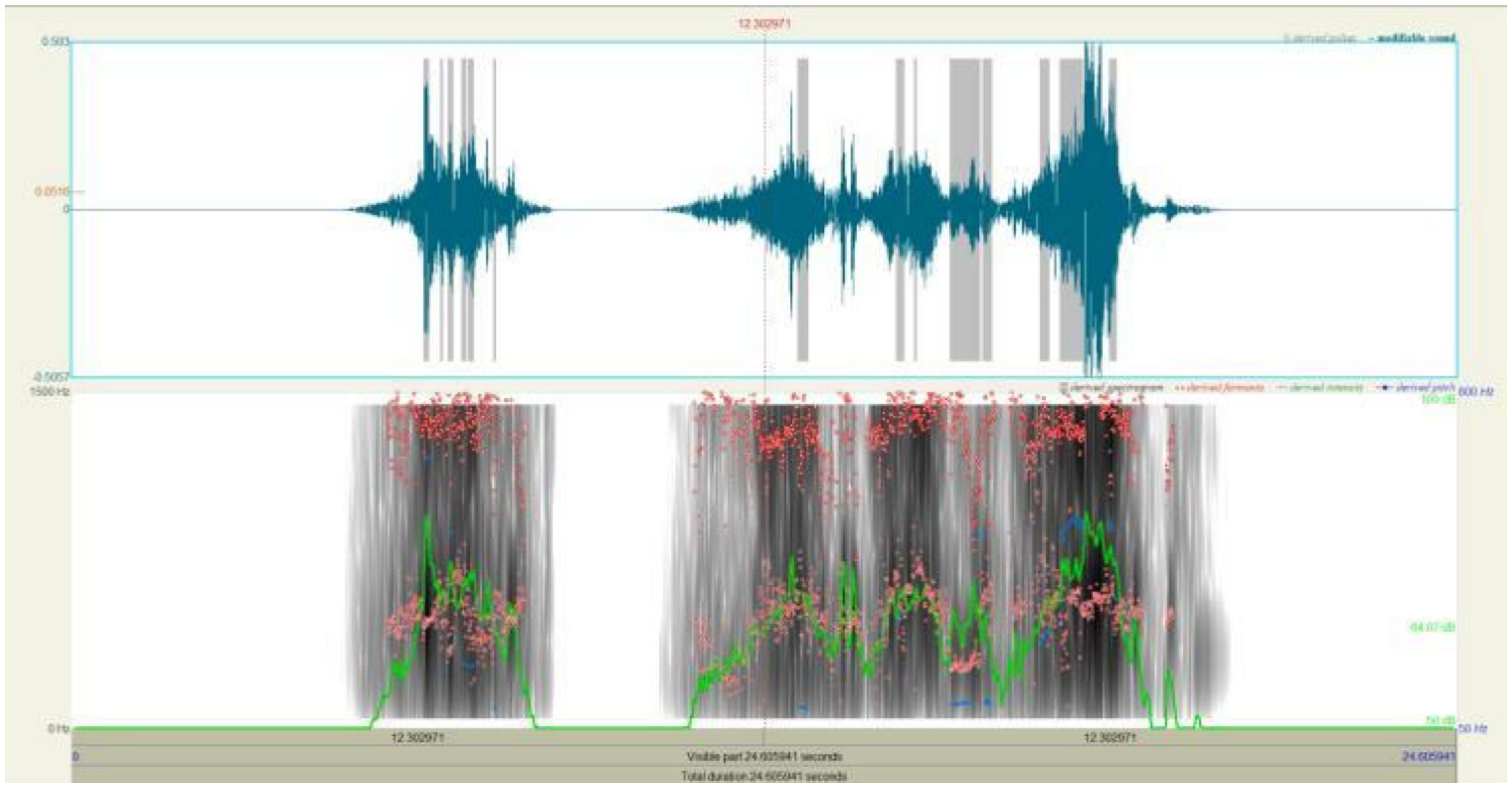

***Figure 8. Acoustic inspection and feature verification in Praat.*** *Screenshot from Praat showing waveform, spectrogram, pitch contour ($F_0$), formant tracks ($F_1$–$F_2$), and intensity envelope for a representative cow vocalization. Visual inspection in Praat was used to verify segmentation boundaries and confirm that selections represented bona fide vocal events rather than background transients. Continuous harmonic structure and expected formant trajectories were*

*checked against known patterns of bovine vocal production, ensuring that annotated clips reflected valid call behavior prior to automated feature extraction and modeling.*

### 4.5 First level annotation

Each segmented clip was subjected to a first-level annotation by two researchers, who assigned multiple labels reflecting call type, emotional context, and confidence, as well as a behavioral summary. Specifically, for each clip the annotator recorded:

- The main category and subcategory according to the scheme defined in Section 2.4, grounded in ethological typologies of vocalizing behavior;
- An emotional context label (distress, pain, anticipation, hunger), reflecting the putative affective state at the time of vocalization;
- A confidence score (1 = low to 10 = high) indicating the annotator's certainty in their labeling;
- A textual description summarizing observable behavioral cues or situational context (e.g. "calf standing near water trough," "cow waiting at feed gate," "walking past milking parlor").

Onset and offset boundaries were further refined within Praat to exploit its high-precision time measurement and pitch/formant display capabilities, allowing annotators to fine-tune temporal limits of calls. The annotation guidelines drew from ethological principles: calls were to be linked to proximate stimuli (e.g. feeding, separation, disturbance) and considered in light of possible ultimate functions (e.g. contact, distress, solicitation). In ambiguous cases or overlapping calls, the label "Unknown" was assigned for later review rather than forcing a classification. Annotators also consulted manual notes and co-recorded video when available to confirm behavioral context (for instance, verifying whether a cow was feeding vs. showing frustration). The output of this stage is a curated set of labeled audio clips, each with category, emotional valence, confidence, and behavioral description, ready for downstream feature extraction and statistical modeling.

In determining the type of call, annotators relied on the spectro-temporal shape (e.g. harmonic stacks, frequency modulation, call duration), amplitude envelope, and context. For example, low-frequency calls with stable harmonics might be assigned to contact or low-arousal categories, whereas calls with abrupt onset, rapid modulation, or high spectral energy could indicate agitation or alarm calls. This practice aligns with literature on cattle vocalization as communicative signals bearing information about motivation or affect (Green et al. 2018) and more broadly on vocal behaviour as an "ethotransmitter" in mammals (Brudzynski 2013). In recent machine-learning work (e.g. (Gavojdian et al. 2024), "BovineTalk"), cattle calls are classified into low-frequency (LF) and high-frequency (HF) types—often mapping LF to close-contact or neutral states and HF to more urgent or negative states—which lends empirical precedent to our annotation categories.

In summary, the preprocessing pipeline we employed—spanning quantitative noise profiling, targeted filtering, professional denoising, careful segmentation, and this rigorous, context-aware annotation—addresses the challenges of noisy barn environments and produces a high-quality, well-documented corpus of vocalizations. This corpus forms a robust foundation for feature extraction and subsequent machine learning or statistical modeling. After preprocessing and initial annotation, we proceeded to compile the dataset through feature extraction, biological interpretation, metadata compilation, and exploratory analysis.

# 5 Dataset Creation

## 5.1 Acoustic feature extraction

Once the segmented clips had undergone first-level annotation, we extracted a comprehensive suite of 24 acoustic features to characterize each vocalization (Table 4). Feature computation was carried out using a combination of Praat, Parselmouth, librosa, and openSMILE, representing both traditional phonetic tools and modern signal-processing libraries. The features spanned temporal, spectral, and cepstral domains, with the aim of capturing both biologically interpretable measures and machine-readable descriptors commonly employed in animal bioacoustic research ((Meen et al. 2015), (Torre et al. 2015)).

- **Temporal metrics** included onset time, offset time, and call duration. Duration has been repeatedly linked to arousal and motivational state: shorter calls are more often associated with neutral or positive contexts, while longer calls typically reflect higher arousal or negative states ((Brudzynski 2013)).

- Signal quality was quantified using **signal-to-noise ratio** (SNR), allowing assessment of how clearly the vocalization emerged from the barn environment. High-SNR clips ensure more reliable feature measurement and downstream modeling.

- **Fundamental frequency** (F0) was measured using Praat's autocorrelation method, with statistics for mean, minimum, and maximum F0 calculated for each clip. Cattle vocalizations generally fall between 50 and 1,250 Hz, with typical averages of 120–180 Hz. High-frequency calls near 150 Hz have been linked to separation distress, whereas low-frequency nasal calls around 80 Hz are associated with close contact and calming social functions ((Watts and Stookey 2000), (Green et al. 2018), (Gavojdian et al. 2024)).

- **Intensity** statistics (minimum and maximum dB) captured variation between quiet and forceful calls. In line with previous findings, higher-intensity calls often indicate urgency or frustration, while lower intensities correspond to calm affiliative contexts ((Watts and Stookey 2000)).

- **Formant frequencies** (F1, F2) were estimated using Praat's Burg method. Formants represent resonances of the vocal tract and provide cues to body size and vocal tract shape ((Fitch and Hauser 2003)). In cattle, F1 and F2 have been documented in ranges between ~ 228 and 3,181 Hz, with average call durations of ~ 1.2 s ((Torre et al. 2015)).

- **Band-level metrics** were extracted using librosa, including bandwidth, RMS energy (mean and standard deviation), spectral centroid, spectral bandwidth, spectral roll-off (85/95), zero-crossing rate (ZCR, mean and standard deviation), and time to peak energy. These descriptors summarize the energy distribution and noisiness of the call. For example, high spectral centroid values correspond to brighter, noisier events (e.g., metallic barn sounds, snorts), whereas lower centroid values are characteristic of harmonic moos.

- **Mel-frequency cepstral coefficients** (MFCCs) (first 13 coefficients: mean and standard deviation) were computed using librosa and openSMILE. MFCCs provide a compact representation of vocal timbre and are widely used in classification of animal calls, including cattle vocalizations((Schrader and Hammerschmidt 1997), (Sattar 2022)).

- Finally, the **voiced rati**o was calculated, representing the proportion of frames classified as voiced versus unvoiced. High-frequency distress calls often exhibit more unvoiced frames due to glottal widening and turbulent airflow, while nasal low-frequency moos are typically fully voiced ((Briefer 2012)).

All features were extracted at a sampling rate of 16 kHz (down-sampled from 44.1/48 kHz), which preserved the relevant bovine vocal bandwidth while reducing computational load.

**Table 4. Acoustic feature set extracted from segmented cattle vocalizations.**

| **Feature** | **Description** | **Extraction Tool / Method** |
|---|---|---|
| Start Time, End Time, Duration | Temporal metrics providing call timing and length. Duration has been linked to arousal and behavioural state: shorter calls often reflect neutral/positive states, longer calls higher arousal. | Praat / Parselmouth |
| Signal-to-Noise Ratio (SNR) | Measures clarity of the call relative to barn noise; ensures robust downstream feature reliability. | Custom script (waveform-based) |
| Fundamental Frequency (F0: mean, min, max) | Pitch contour statistics; cattle vocal range ~ 50–1250 Hz. High frequencies linked to distress/separation, low nasal calls (~80 Hz) to close contact. | Praat autocorrelation / Parselmouth |
| Intensity (min, max, mean dB) | Energy levels of the vocalization; high intensity = urgency/frustration, low intensity = calm contact. | Praat / Parselmouth |
| Formants (F1, F2) | Resonant vocal tract frequencies; provide cues to body size and vocal configuration. | Praat Burg method |
| Spectral Centroid | Centre of mass of spectrum; high values = bright/noisy, low values = harmonic moos. | librosa |
| Spectral Bandwidth | Spread of spectral energy around the centroid. | librosa |
| Spectral Roll-off (85%, 95%) | Frequency below which 85% or 95% of spectral energy is contained. | librosa |
| Zero Crossing Rate (mean, std) | Rate of waveform sign changes; indicates noisiness/aperiodicity. | librosa |
| RMS Energy (mean, std) | Root mean square energy; reflects call strength and stability. | librosa |

| Feature | Description | Extraction Tool / Method |
|---|---|---|
| Time to Peak Energy | Time taken for a call to reach maximum energy; a dynamic marker of urgency. | librosa / scipy |
| Mel Frequency Cepstral Coefficients (MFCCs 1–13: mean, std) | Cepstral features capturing vocal timbre; widely used in vocal classification. | librosa / openSMILE |
| Voiced Ratio | Proportion of voiced vs. unvoiced frames; distress calls often more voiceless, nasal moos nearly fully voiced. | Praat / Parselmouth |

Beyond numerical computation, these features hold biological meaning, which we interpret in the next subsection.

### 5.2 Biological interpretation of Acoustic features

The selected features were chosen not only for their statistical utility in machine learning, but also for their biological interpretability. This dual emphasis ensures that computational models remain grounded in the mechanisms of cattle vocal production and their ethological significance.

- **Fundamental frequency** (F0) is primarily determined by the length, tension, and mass of the vocal folds ((Brudzynski 2010)). Increases in F0 are commonly associated with heightened arousal, separation distress, or estrus, while lower F0 is characteristic of calm affiliative calls ((Green et al. 2018), (Röttgen et al. 2018)). In our dataset, calls labelled Estrus_Call and High_Frequency_Distress exhibited both elevated maximum F0 and greater F0 variability, consistent with reports that female cattle emit higher-pitched calls during estrus or when separated from their calves. Conversely, low-frequency nasal moos, typically around 80–120 Hz, were more often associated with affiliative or contact-seeking contexts, echoing findings from maternal–offspring communication studies ((Torre et al. 2015)).

- **Formant frequencies** (F1, F2) reflect vocal tract resonances and convey information about articulatory configuration and body size ((Fitch and Hauser 2003)). Lower formant values are linked to mouth opening and longer vocal tract length, while higher formants reflect tongue and lip positioning. In our data, Water_Slurping_Sounds displayed broad bandwidths and high F1–F2 separation, capturing their noisy, non-harmonic structure, whereas harmonic moos showed tighter clustering of F1 and F2 bands. This observation is consistent with earlier descriptions of formant dynamics in cattle vocalizations ((Watts and Stookey 2000)).

- **Energy-based measures** (RMS energy, intensity, and time to peak) further captured the dynamic and affective force of vocalizations. High RMS energy and short time-to-peak values were prominent in Frustration_Calls and Aggressive_Bellows, reflecting abrupt, high-force emissions. By contrast, Mother_Separation_Calls were lower in intensity but longer in duration, representing persistent vocal efforts at lower force levels, in line with observations of separation calls in cow–calf pairs ((Green et al. 2019)).

- **Spectral descriptors** provided a robust means of distinguishing harmonic from noisy events. Spectral centroid and roll-off values differentiated stable harmonic moos from broadband, non-vocal sounds such as sneezes, coughs, or burps. These non-vocal events showed elevated centroid and zero-crossing rate values, indicating their noisy and aperiodic character ((Meen et al. 2015)).
- Finally, **Mel-frequency cepstral coefficients** (MFCCs) encoded global spectral shape and proved particularly useful for distinguishing subtle differences between similar categories such as Feed_Anticipation_Call and Feeding_Frustration_Call. This aligns with recent applications of MFCCs in livestock monitoring, where cepstral analysis has been shown to detect estrus events with accuracies exceeding 90% ((Sattar 2022), (Gavojdian et al. 2024)).

Taken together, these feature–behavior associations reinforce that the dataset is not only suitable for computational modeling, but also biologically meaningful. By anchoring machine-readable features to established ethological frameworks (Tinbergen's four questions;(Dugatkin 2020)), the approach ensures that downstream classification and prediction retain relevance to animal welfare science and practical dairy farm monitoring.

**Table 5. Mapping of acoustic features to biological interpretation and representative call types in the dataset.**

| **Acoustic Feature** | **Biological Interpretation** | **Example Call Types** |
|---|---|---|
| Fundamental Frequency (F0: mean, min, max) | Determined by vocal fold tension, length, and mass; high F0 reflects arousal, distress, or estrus, while low F0 indicates calm affiliative contact. | *Estrus_Call*, *High_Frequency_Distress*, low-frequency moos (contact) |
| Formant Frequencies (F1, F2) | Resonances of the vocal tract linked to mouth opening, tongue/lip position, and body-size cues; larger F1–F2 separation often accompanies noisier or less harmonic structure. | *Water_Slurping_Sounds*, harmonic moo (stable formants) |
| Duration & Timing (Start, End, Duration) | Persistence of calling; shorter calls often reflect neutral/positive states, whereas longer calls are associated with higher arousal or separation. | *Mother_Separation_Call* (long, low intensity); *Feed_Anticipation_Call* (short bursts) |
| Energy Measures (RMS, Intensity, Time to Peak) | Reflect call forcefulness and emotional valence; high RMS and fast time-to-peak indicate urgency, while low intensity with long duration suggests persistent, subdued calls. | *Aggressive_Bellow*, *Frustration_Call*; calf contact moo (low intensity) |
| Spectral Centroid, Bandwidth, Roll-off, Zero-Crossing Rate (ZCR) | Differentiate harmonic vs. noisy events; high centroid/ZCR imply noisy or aperiodic content, low centroid indicates harmonic structure. | Sneezes, burps (high centroid, high ZCR); harmonic moo (low centroid, low ZCR) |
| Mel-Frequency Cepstral | Capture global spectral shape and | *Feed_Anticipation_Call* |

| Acoustic Feature | Biological Interpretation | Example Call Types |
|---|---|---|
| Coefficients (MFCCs) | timbre; useful for subtle distinctions between similar call types, with MFCC-based estrus detection reported at > 90% accuracy. | vs. *Feeding_Frustration_Call* |
| Voiced Ratio | Proportion of voiced vs. unvoiced frames; distress calls tend to include more unvoiced segments, whereas nasal moos are almost fully voiced. | High-frequency distress calls (more unvoiced); nasal moos (fully voiced) |

To ensure each feature and annotation is transparent and reusable, we compiled a structured metadata schema.

### 5.3 Metadata compilation and schema

To ensure that each audio clip could be unambiguously identified, contextualized, and reused for further research, we developed a comprehensive metadata schema to accompany the curated dataset (Table 6). The schema integrates identifiers, contextual descriptors, acoustic parameters, behavioral annotations, and environmental measures, thereby aligning with best practices for animal genomics and behavioral data curation as outlined by the FAANG consortium ((Harrison et al. 2018)) and FAIR data principles.

Metadata fields were structured into five categories:

1. **File identifiers:** Each clip is assigned a unique filename following a structured convention that encodes farm ID, recording zone, date, time, microphone ID, and provisional call type. Additional fields include the original recording filename and precise date–time stamps.

2. **Contextual information:** These fields capture the recording environment and instrumentation, including farm identifier, barn zone (e.g., feeding area, water station, resting area, milking parlor), microphone model, recorder type, and mic placement context. Such metadata ensure interpretability across heterogeneous barn environments ((Meen et al. 2015)).

3. **Acoustic features:** The full set of 24 acoustic features described in Section 5.1 — including duration, F0 statistics, formant values, intensity measures, spectral centroid, bandwidth, roll-off, RMS energy, zero-crossing rate, MFCC statistics, voiced ratio, and time-to-peak energy—are included. Storing these features directly in the metadata table enables rapid subsetting and analysis without rerunning extraction pipelines.

4. **Behavioral annotations:** Each clip is linked to the annotation schema described in Section 4.5. Fields include the main category and subcategory (e.g., Feeding_Anticipation_Call, Mother_Separation_Call, Burp), emotional context (discomfort, pain, hunger, thirst), annotator confidence score (1–10), and free-text description summarizing behavioral cues (e.g., "calf standing near water trough," "cow waiting at feed gate"). This dual annotation—structured categories plus free-text notes—enables both quantitative and qualitative analysis.

5. **Environmental parameters:** To document recording conditions and preprocessing steps, metadata also include the low and high frequency cut-offs from the band-pass filter (Section 4.2), effective bandwidth, signal-to-noise ratio, and microphone gain settings where available. These parameters are critical for reproducibility, given the variability of barn acoustic environments ((Alsina-Pagès et al. 2021)).

The metadata schema thus supports Findable, Accessible, Interoperable, and Reusable (FAIR) data practices by providing clear identifiers, structured descriptors, and machine-readable acoustic metrics. To illustrate, Tables 7 & 8 presents three representative metadata records: a Feed_Anticipation_Call recorded in the feeding zone, a Mother_Separation_Call in the resting area, and a Burp in the feeding zone. These examples demonstrate how behavioral categories, recording contexts, and acoustic profiles (e.g., differences in duration, F0 maxima, intensity, and formant dispersion) are transparently documented.

By compiling both biological and technical descriptors, the metadata not only enable efficient subsetting of the corpus (e.g., by behavior, environment, or acoustic property), but also ensure long-term reproducibility and interoperability with other livestock bioacoustic datasets.

**Table 6. Metadata schema accompanying each segmented clip.**

| Field | Description | Data Type | Example |
|---|---|---|---|
| File Name | Unique identifier for the clip, following structured naming convention | String | Farm1_Drinking_2025-05-05_10-45_Mic2_Feed_Anticipation.wav |
| Original File | Parent recording file name | String | Farm1_Day1_WaterStation.wav |
| Date | Date of recording | Date (YYYY–MM–DD) | 2025-05-06 |
| Time | Time of recording | Time (HH:MM:SS) | 10:45:32 |
| Farm ID | Unique identifier for farm | Integer / String | Farm1 |
| Barn Zone | Location within barn (e.g., feeding, water, resting, milking area) | String | Feeding Zone |
| Microphone Model | Microphone model used | String | Rode NTG2 |
| Recorder | Recorder type | String | Zoom H4n Pro |
| Mic Placement Context | Placement or mounting details | String | Above feed trough |
| Start Time (s) | Start time of clip within original recording | Float (seconds) | 938.03 |

| Field | Description | Data Type | Example |
|---|---|---|---|
| End Time (s) | End time of clip within original recording | Float (seconds) | 963.84 |
| Duration (s) | Call duration | Float (seconds) | 25.81 |
| Fundamental Frequency (F0) | Mean, min, and max pitch values | Float (Hz) | Mean: 150 Hz; Max: 310 Hz |
| Formants (F1, F2) | First and second formant frequencies | Float (Hz) | F1 = 320 Hz, F2 = 1180 Hz |
| Energy Metrics | Intensity (min, max, mean dB); RMS energy | Float (dB) | Min = 55 dB, Max = 70 dB |
| Spectral Features | Centroid, bandwidth, roll-off, zero crossing rate, time to peak | Float / Derived | Centroid = 1120 Hz; ZCR = 0.08 |
| MFCCs (1–13) | Mean and standard deviation of MFCC coefficients | Array (floats) | [12.4, 9.6, …] |
| Voiced Ratio | Proportion of frames classified as voiced vs. unvoiced | Float (%) | 92% voiced |
| Main Category | Behavioral category (from annotation scheme) | String | Feeding and Hunger Related |
| Sub Category | Specific vocal type | String | Feed_Anticipation_Call |
| Emotional Context | Affective state inferred from annotation | String | Positive |
| Confidence Score | Annotator certainty level (1 = low, 3 = high) | Integer (1–3) | 3 |
| Description | Free-text summary of behavior/context | String | Cow waiting at feed gate |
| Low Frequency Bound | Lower cutoff of applied band-pass filter | Float (Hz) | 50 Hz |
| High Frequency Bound | Upper cutoff of applied band-pass filter | Float (Hz) | 1800 Hz |
| Bandwidth | Effective bandwidth of call | Float (Hz) | 1750 Hz |
| Signal-to-Noise Ratio | Call clarity relative to background | Float (dB) | 15.3 dB |

**Table 7. Rep. metadata records (A): identifiers and context.**

| File Name | Barn Zone | Category | Description |
|---|---|---|---|
| Feeding_Hunger_Empty_Feeder_Call_ | Feeding | Empty_Feeder | Complaint at empty |

| File Name | Barn Zone | Category | Description |
|---|---|---|---|
| 34s | | _Call | feeder |
| Maternal_Calf_Separation_Mothercow _Call_13s | Resting | Mother_Separa tion_Call | Loud, urgent call for calf |
| Non_Vocal_Chewing_Burping_Sound _07s | Feeding | Burp | Chewing and cud sounds |

**Table 8. Rep. metadata records (B): acoustic measurements.**

| (s) | (Hz) | (dB) | (Hz) | (Hz) | File Name (short) |
|---|---|---|---|---|---|
| 34.22 | 368.0 | 26.65 | 650.67 | 1653.99 | Empty_Feeder_Call_34s |
| 13.00 | 615.3 | 60.44 | 1079.64 | 1913.06 | Mother_Separation_Call |
| 6.84 | 29.7 | 28.89 | 740.71 | 1951.26 | Burping_Sound_07s |

### 5.4 Class distribution, imbalance and augmentation

The final curated dataset comprises 569 audio clips spanning 48 behavioral classes. Each clip is a denoised vocalization or nonvocal sound with precise onset and offset boundaries. The mean duration is ~ 21 s (median ~ 13.8 s; range 2.8–445 s), with ~ 75 % of clips shorter than 21 s. The three most frequent classes—Estrus_Call (117 clips), Feed_Anticipation_Call (113 clips), and Breathing_Respiratory_Sounds (76 clips)—account for 53 % of the dataset, whereas 21 classes contain fewer than 10 clips each. This pronounced class imbalance reflects the natural occurrence of certain behaviours and the rarity of others. Compared with earlier bovine vocalisation datasets (e.g., 1,144 calls from 20 isolated cows (Gavojdian et al. 2024), or 290 calls across four physiological states (Yoshihara and Oya 2021)), our corpus is broader in scope, encompassing multiple farms, barn zones, high- and low-frequency calls, and nonvocal events. Skewed distributions are common in animal bioacoustics ((Kahl et al. 2021)) and present well-documented challenges for machine learning models, which may overfit to frequent classes and underperform on rare but ethologically meaningful categories ((Johnson and Khoshgoftaar 2019)).

To mitigate this imbalance, we adopted a two-pronged approach combining data augmentation with class-balanced sampling:

1. **Data augmentation** of the training set. Each clip was subjected to biologically informed perturbations, including random time-stretching (0.8–1.2 × ), pitch shifting ( ±2 semitones), Gaussian noise addition (SNR ≥ 20 dB), and gain adjustment (±6 dB). These augmentations simulate natural variability in vocal production and recording context, such as differences in caller distance, vocal effort, or microphone orientation. Importantly, augmentation parameters were constrained to remain within biologically plausible ranges: pitch shifts avoided unrealistic F0 values, while time-stretching was limited to ±20% to preserve temporal dynamics of vocal events ((Briefer 2012), (Ganchev et al. 2005)).

2. **Class-balanced sampling and oversampling**. During training, mini-batches were constructed with equal representation from each class. For classes with < 10 clips, augmented data were used to oversample minority categories. Extremely rare classes (< 3 clips) were excluded from training but retained in the metadata corpus for future research, thereby avoiding overfitting while ensuring transparency of available data.

After augmentation, the training set expanded from 569 to approximately 2,900 clips, providing a sufficiently balanced dataset for model training. The validation and test sets retained only the original, non-augmented clips to ensure unbiased evaluation. Table 9 presents the number of clips per class before and after augmentation, demonstrating how augmentation improved distributional balance.

By combining augmentation and balanced sampling, the approach reduced class bias and improved the capacity of models to generalize across both frequent calls (e.g., feeding anticipation, respiratory sounds) and rare but behaviorally significant calls (e.g., social recognition, separation). This methodology reflects best practice in animal bioacoustics, where careful augmentation helps capture the ecological variability of vocal signals without compromising their biological validity.

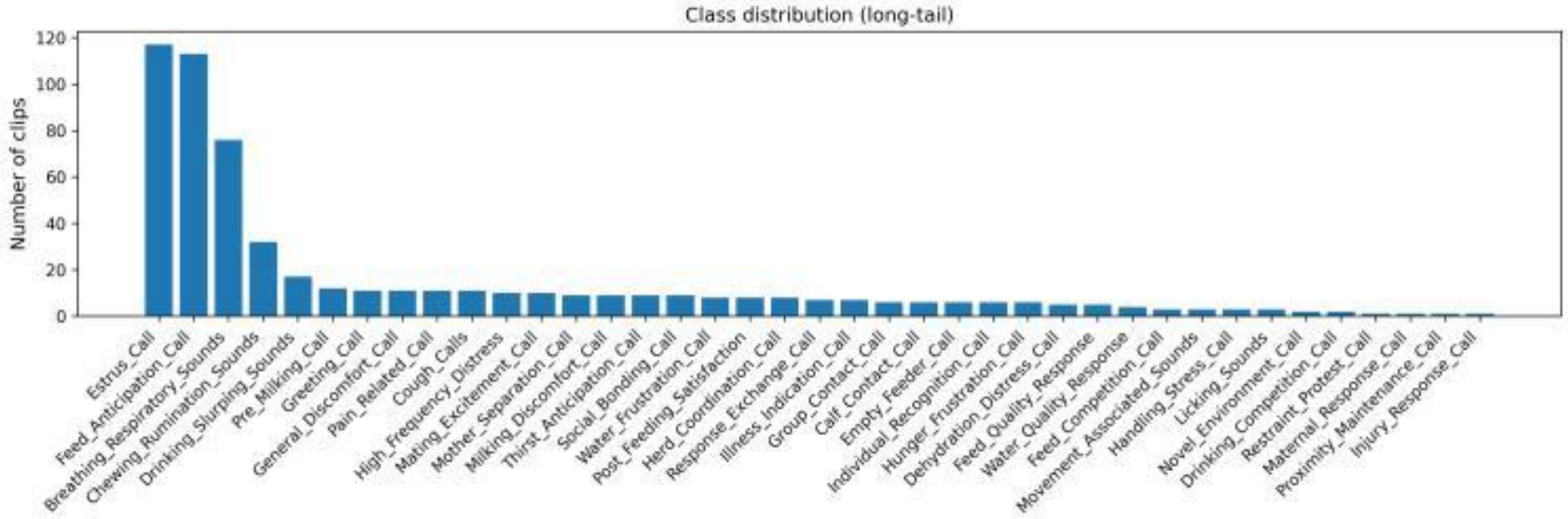

***Figure 9. Overall class distribution of cow vocalizations.*** *Long-tail distribution of all annotated classes in the dataset. Each bar represents the number of audio clips within a behavioral or acoustic category. A pronounced imbalance is evident, with a few highly represented classes (e.g., estrus and anticipation calls) and many rare categories such as response exchange or distress calls, reflecting the natural skew of on-farm acoustic events.*

**Table 9. Representative examples of clip counts before and after augmentation.**

| Class | Before Augmentation | After Augmentation |
|---|---|---|
| Estrus_Call | 117 | 81 |
| Feed_Anticipation_Call | 113 | 81 |
| Breathing_Respiratory_Sounds | 76 | 81 |
| Chewing_Rumination_Sounds | 32 | 81 |
| Drinking_Slurping_Sounds | 17 | 81 |
| Pre_Milking_Call | 12 | 81 |
| Cough_Calls | 11 | 81 |
| Pain_Related_Call | 11 | 81 |

| Class | Before Augmentation | After Augmentation |
|---|---|---|
| Maternal_Response_Call | 1 | Dropped |
| Restraint_Protest_Call | 1 | Dropped |
| Proximity_Maintenance_Call | 1 | Dropped |

### 5.5 Preliminary feature analysis

To assess the discriminative power of the extracted features, we conducted exploratory analyses of temporal, spectral, and energy-based parameters across the dataset. Clip duration (Fig. 10 exhibited a heavy-tailed distribution: the majority of clips were shorter than 20 s, although some extended beyond 200 s; the 90th percentile was ~ 60 s. This skewed distribution reflects the behavioral variability of cattle vocalizations, where most calls are short, context-bound signals, while prolonged calls occur during high-arousal contexts such as estrus or maternal separation ((Röttgen et al. 2018);(Green et al. 2018)).

Pitch statistics (F0) (Fig. 13) varied significantly across call categories. Non vocal sounds like Breathing, Chewing and Mother_Separation_Calls showed low mean F0 values (~ 100 Hz), consistent with affiliative and maternal communication, while High_Frequency_Distress, Feed_Anticipation_Calls and Estrus_Calls exhibited mean F0 values above 150 Hz, in line with literature linking high-frequency vocalizations to heightened arousal and distress ((Harrison et al. 2018), (Torre et al. 2015)). These observations corroborate ethological studies describing F0 as a reliable correlate of motivational state and affect in cattle.

Energy measures (Fig. 14) also provided discriminative value. Non-vocal expulsive sounds such as Breaths, Sneezes, and Coughs were characterized by high RMS energy and elevated zero-crossing rates, consistent with their noisy, aperiodic spectral structure. In contrast, harmonic moos displayed lower RMS values and more stable zero-crossing patterns, reflecting their tonal and harmonic nature ((Meen et al. 2015)).

To statistically evaluate feature variability, we applied non-parametric Kruskal–Wallis tests on key features (clip duration, mean F0, spectral centroid, and RMS energy) across the twelve largest classes ($n \geq 10$). Results revealed highly significant differences ($p < 0.001$) in F0 and spectral centroid across classes, while duration differences were less pronounced. For instance, median spectral centroid values for Water_Slurping_Sounds were more than double those of Mother_Separation_Calls, reflecting the broadband, noisy profile of slurping compared with the harmonic structure of vocal calls.

Feature distributions were visualized using violin plots (Fig. 12, 13, 14 ), which highlighted class-specific signatures: Estrus and Feed_Anticipation_Calls clustered at higher F0 ranges, Moos and Mother_Separation_Calls at lower F0 ranges with narrow spectral spread, and non-vocal events (Burps, Sneezes, Slurps) exhibited wider energy and spectral distributions. Together, these findings demonstrate that the extracted features capture meaningful behavioral variation and provide a robust foundation for downstream machine learning classification. Together, these steps resulted in a curated, feature-rich dataset. The following discussion examines its broader significance for AI and animal welfare research, acknowledges current limitations, and highlights future opportunities.

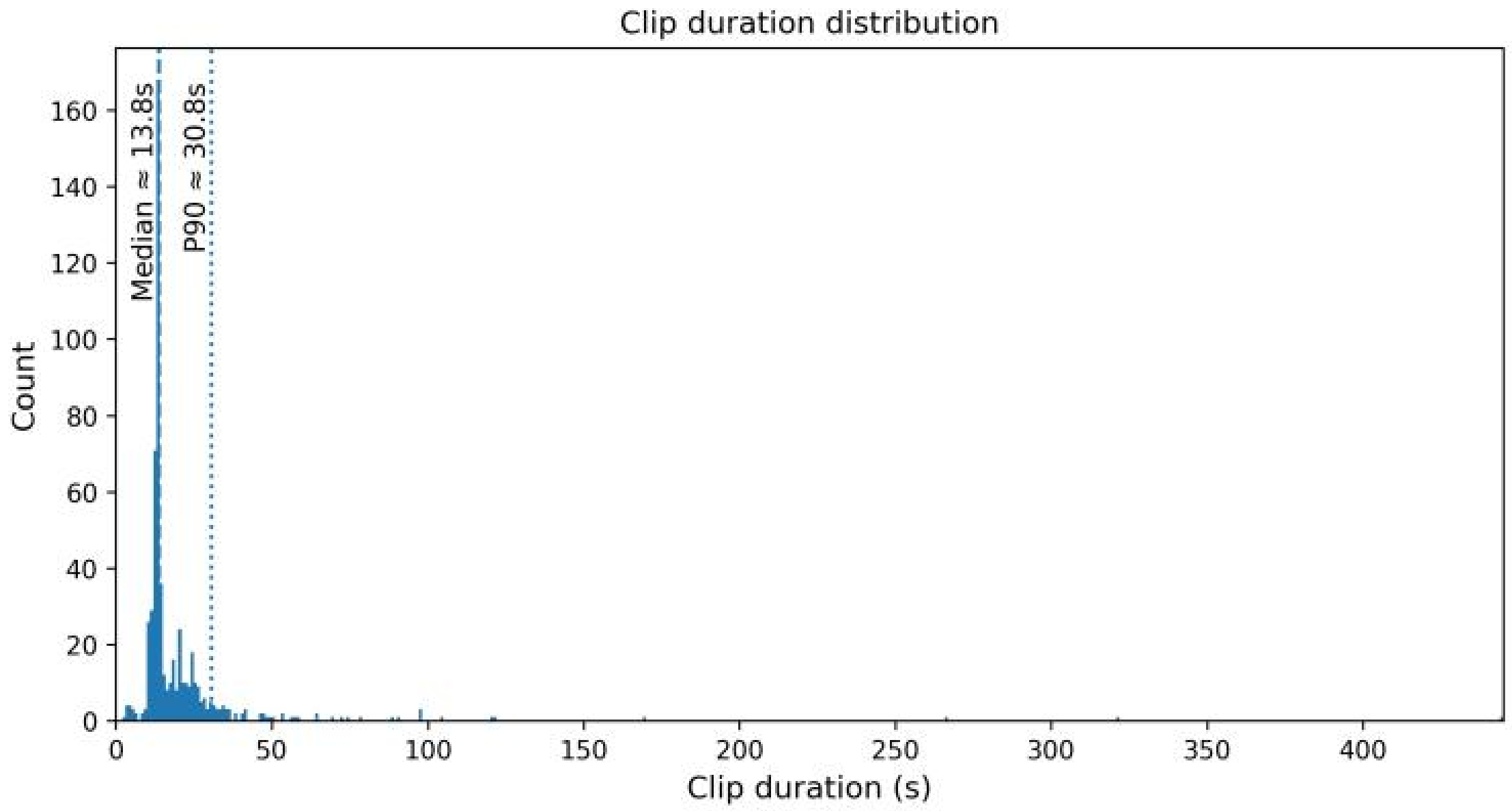

***Figure 10. Clip duration distribution across the dataset.*** *Histogram showing the temporal distribution of clip lengths for all annotated vocalizations. Most clips are short (under 20 s), while a small number extend beyond one minute, producing a heavy-tailed pattern. Dashed and dotted lines mark the median and 90th-percentile durations, respectively, highlighting the variability in call length across behavioral contexts.*

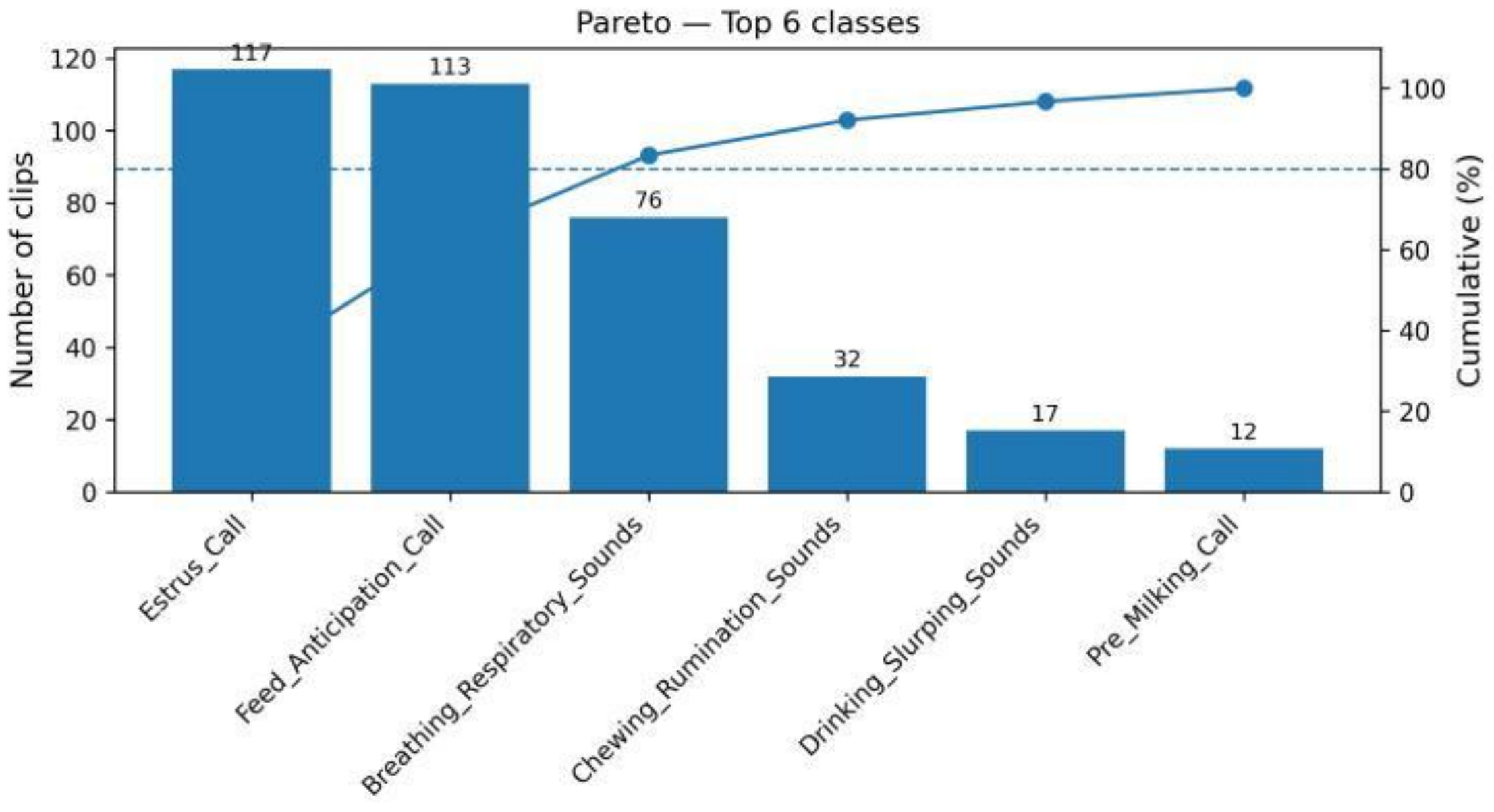

***Figure 11. Pareto chart of top six vocalization classes.*** *Distribution of the most frequent vocalization categories in the dataset, displayed as a Pareto chart. Bars indicate the number of clips per class, and the overlaid line shows the cumulative proportion of all clips. The six dominant classes together account for the majority of recorded samples, illustrating the long-tailed nature of class occurrence and guiding model evaluation toward balanced and minority-aware analysis.*

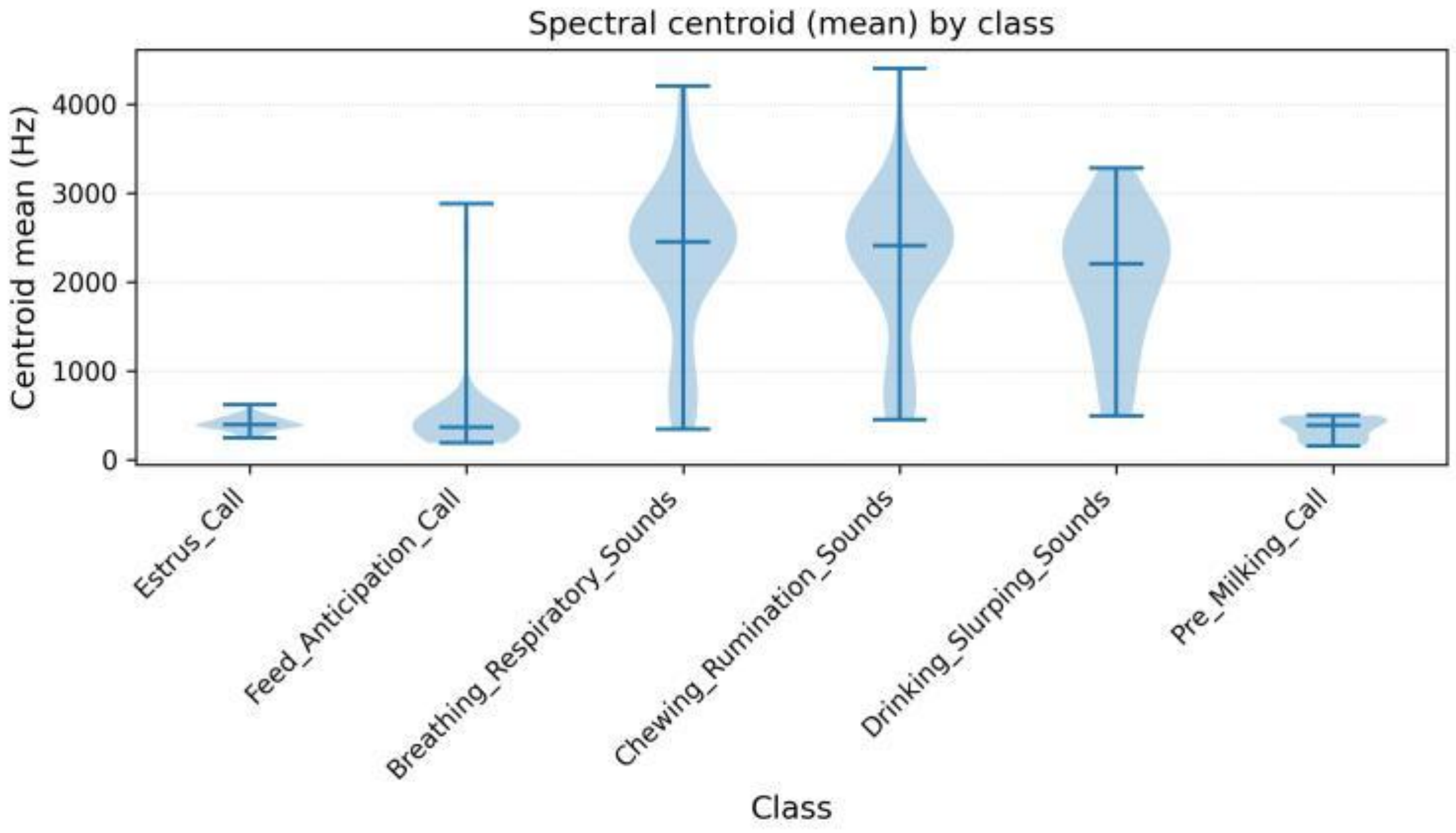

***Figure 12. Violin plot of spectral centroid (mean) by class.*** *Distribution of mean spectral centroid values for the top six classes. The spectral centroid describes the "brightness" of a sound, with higher values indicating stronger high-frequency components. The plot shows clear class-wise differences in spectral coloration—harmonic moos cluster at lower centroids, whereas short impulsive events (e.g., coughs, sneezes) show broader, higher-frequency spectra.*

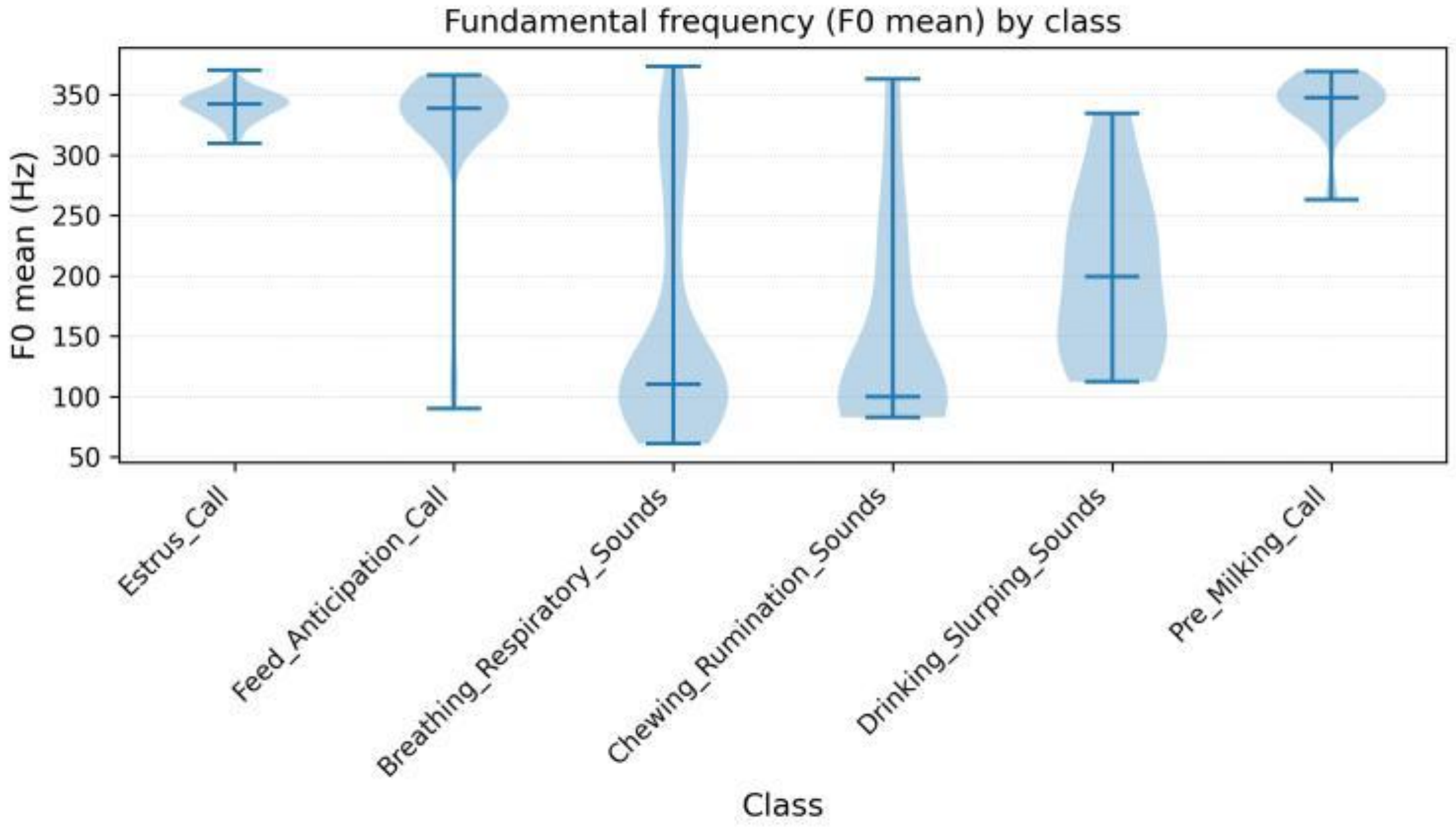

***Figure 13. Violin plot of fundamental frequency ($F_0$ mean) by class.*** *Variation in mean fundamental frequency ($F_0$) across the top six classes. Each violin depicts the distribution of $F_0$ values within a class, with the median shown by a central line. Lower-frequency ranges*

*correspond to low-arousal or contact calls, while higher-frequency calls are associated with heightened arousal or distress, consistent with previous reports in bovine vocal studies.*

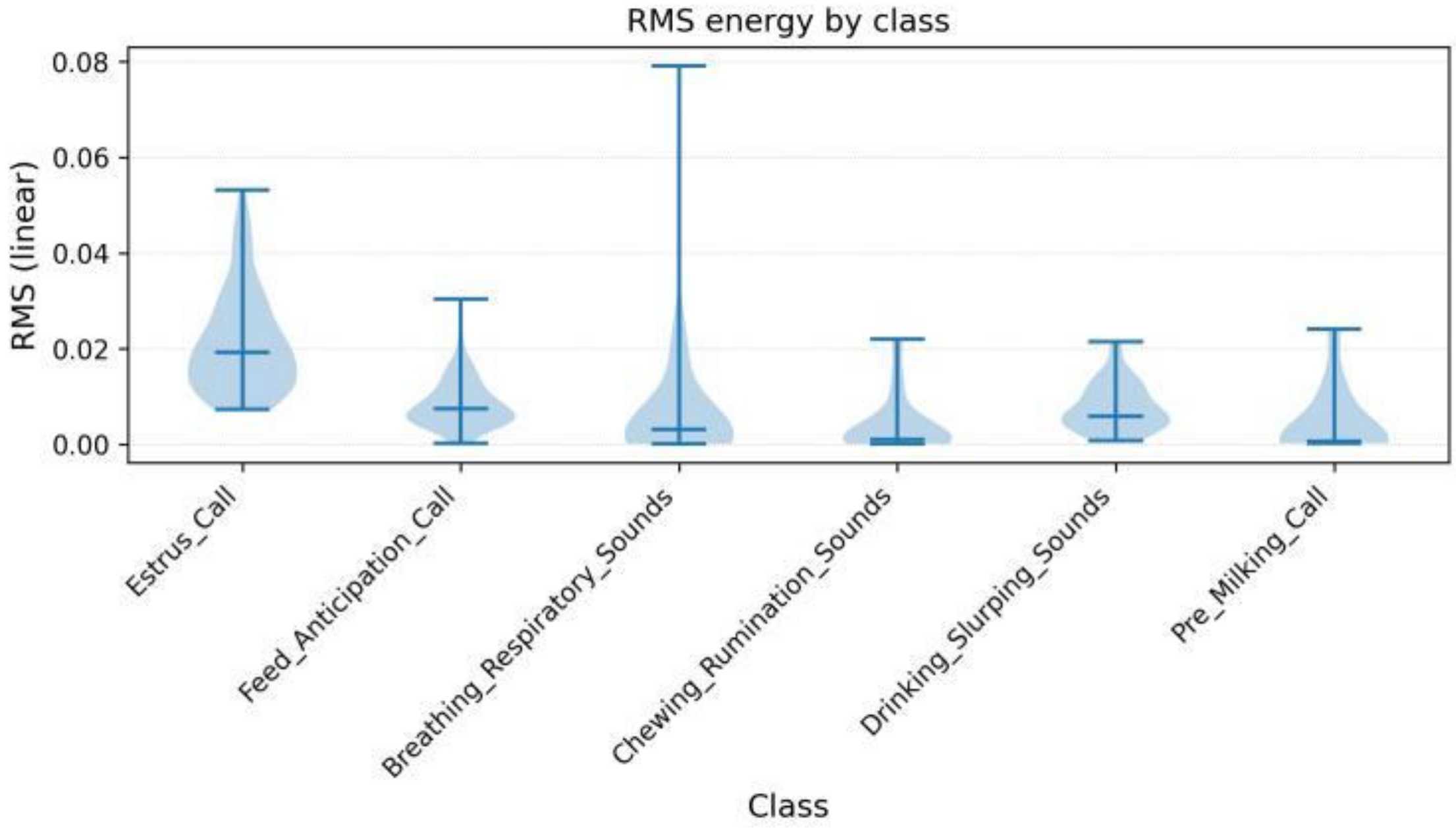

***Figure 14. Violin plot of RMS energy by class.*** *Comparison of average root-mean-square (RMS) energy among the top six vocalization classes. The RMS measure reflects overall signal intensity and vocal effort. Classes such as coughs and burps exhibit higher energy levels due to their abrupt, broadband nature, whereas harmonic moos and contact calls show lower, more stable energy envelopes.*

# 6 Discussion

## 6.1 Significance for AI and Big Data

The dataset presented in this study contributes directly to advancing AI and data-driven approaches in animal bioacoustics. Unlike prior collections that were limited in size or scope, this corpus offers both scale and diversity, providing 569 annotated clips (expanded to ~ 2,900 after augmentation for modelling) across 48 behavioural classes. Such breadth is crucial for machine learning applications, where model performance depends on exposure to both common and rare events. By incorporating high-arousal calls (e.g., estrus, distress), affiliative calls (e.g., contact, maternal), and non-vocal sounds (e.g., breathing, burps), the dataset enables algorithms to learn the acoustic signatures of a wide behavioural spectrum rather than a narrow subset of conditions.

Equally important is the dataset's ecological realism. Machine learning models trained on clean, noise-free recordings often fail when deployed in real barn environments. By retaining authentic background noise and overlapping acoustic events, this resource ensures that computational models developed from it are robust to the challenges of deployment. The multimicrophone and multimodal setup adds further depth, allowing researchers to explore cross-validation across equipment types and zones, or to align acoustic features with behavioural context captured in video. This design makes the dataset valuable not only for supervised classification but also for self-supervised learning, representation learning, and transfer learning, areas where large, heterogeneous datasets are particularly impactful.

Finally, the integration of FAIR-compliant metadata and standardised acoustic features (Praat, librosa, openSMILE) ensures that the dataset is interoperable with wider AI research ecosystems. Researchers can directly apply established pipelines for feature selection, dimensionality reduction, or deep learning input preparation, reducing barriers to reproducibility. In this way, the dataset bridges the gap between animal behaviour research and contemporary AI methodologies, situating bovine vocalization analysis firmly within the domain of big data science.

### 6.2 Limitations and Challenges

While comprehensive, the dataset is not without limitations. The most prominent challenge lies in class imbalance, with a small number of categories such as estrus and feeding anticipation dominating the corpus, while rare events like drinking competition calls or novel environment response calls are represented by only a few clips. This long-tailed distribution mirrors behavioural ecology but complicates model training, as classifiers may overfit frequent categories while neglecting rare yet ethologically significant ones. Augmentation strategies can mitigate this bias, but they cannot fully replace naturally occurring data.

Another limitation arises from the subjectivity of manual annotation. Although rigorous protocols and cross-validation by multiple annotators were employed, certain call types are inherently ambiguous, especially when overlapping with background noise or occurring in complex social contexts. Labels such as “frustration” or “distress” are based on behavioural inference, which, while grounded in ethology, cannot capture internal emotional states with absolute certainty. This highlights the need for future datasets to integrate multimodal physiological or sensor-based validation to strengthen label reliability. Manual segmentation and annotation, while ensuring high fidelity, remain time-intensive and may limit scalability without semi-automated approaches. Future work integrating automated detectors or active learning frameworks could alleviate this bottleneck while preserving annotation accuracy.

The acoustic environment itself also posed challenges. Commercial barns are characterised by persistent mechanical noise and overlapping vocal activity, which, although essential for ecological realism, reduce signal clarity. Even after careful denoising and filtering, residual interference remains in some clips. While this realism enhances deployment value, it also increases computational demands for segmentation, feature extraction, and classification. Researchers applying advanced models to this dataset should therefore be aware that noise robustness remains an open problem.

Finally, although the dataset is large by livestock bioacoustics standards, it remains modest compared to big data benchmarks in other AI fields such as speech recognition or computer vision. Expanding the temporal coverage (e.g., across seasons, farms, and breeds) and enlarging the sample size would further strengthen its generalisability and enable the training of more complex deep learning architectures.

### 6.3 Outlook and Future Directions

The current dataset establishes a foundation, but it also opens several avenues for expansion and methodological innovation. One promising direction is the incorporation of longitudinal recordings that capture vocal behaviour across different seasons, management practices, and life stages. Extending coverage beyond three farms and including diverse breeds would improve representativeness and allow for comparative studies across genetic and environmental contexts.

Another opportunity lies in multimodal integration. While this dataset already links audio to video and manual annotations, future work could align vocal data with physiological markers (e.g., heart rate, cortisol levels, rumination sensors) to provide multi-layered evidence of welfare states. Such integration would reduce ambiguity in behavioural labels and strengthen the interpretability of acoustic indicators.

From a computational perspective, the dataset is particularly well-suited for exploring emerging AI paradigms. Large, heterogeneous acoustic corpora are valuable for training self-supervised models that learn general-purpose representations before fine-tuning for specific tasks such as estrus detection, welfare monitoring, or individual identification. Similarly, transfer learning from bovine vocalizations to related species—or from human speech models to livestock contexts—presents opportunities for cross-domain innovation.

Finally, the dataset highlights the importance of open, standardised resources in agricultural AI. By adhering to FAIR principles, this work contributes to a growing movement toward reproducible, community-driven datasets in animal science. Establishing shared benchmarks for livestock bioacoustics, similar to ImageNet or LibriSpeech in computer vision and speech research, would accelerate progress by enabling systematic comparisons of models and fostering collaborative development. The resource introduced here can serve as an early step in that direction, encouraging both the scaling of future datasets and the refinement of analytical tools tailored to animal vocal behaviour. These considerations set the stage for our conclusion, where we summarise the dataset's contributions and its potential role in advancing digital agriculture and big data applications in animal science.

## 7 Conclusion

In this work, we introduced one of the most comprehensive bovine vocalization datasets assembled to date, integrating 569 curated original clips across 48 behavioural classes recorded in authentic barn environments. By combining multi-microphone audio capture, complementary video observations, and detailed ethology-driven annotations, the dataset provides an ecologically valid resource for advancing both animal behaviour research and computational modelling. Preprocessing steps ensured that clips were denoised, segmented, and paired with rich metadata, while feature extraction pipelines generated interpretable acoustic descriptors aligned with welfare science.

The dataset makes three key contributions. First, it broadens the empirical foundation for studying cattle vocal behaviour beyond the constraints of controlled laboratory recordings, embracing the acoustic complexity of real farm settings. Second, it provides a reproducible, FAIR-compliant framework with transparent metadata and feature definitions, positioning bovine bioacoustics within the wider ecosystem of big data research. Third, it offers a benchmark corpus for the development and testing of AI methods, from supervised classification to emerging approaches such as self-supervised representation learning.

Taken together, these contributions establish a foundation for non-invasive, data-driven approaches to animal welfare monitoring and precision livestock management. While challenges remain in scaling, class balance, and multimodal integration, this dataset represents a critical step toward creating the robust, reproducible resources needed to give cattle a “digital voice” in future smart farming systems.

## Conflict of Interest Statement

The author declares that the research was conducted in the absence of any commercial or financial relationships that could be construed as a potential conflict of interest. The authors declared that Professor Suresh Neethirajan was an editorial board member of Frontiers, at the time of submission. This had no impact on the peer review process and the final decision.

## Author Contributions

MK: Investigation, Data Collection, Writing - Original Draft, SN: Conceptualization, Funding acquisition, Investigation, Project administration, Resources, Writing - review & editing

## Funding

The author(s) declare that financial support was received for the research and/or publication of this article. The authors thank the Natural Sciences and Engineering Research Council of Canada (NSERC), Nova Scotia Department of Agriculture, Mitacs Canada, and the New Brunswick Department of Agriculture, Aquaculture and Fisheries for funding this study.

## Acknowledgments

The authors sincerely thank the Dairy Farmers of Nova Scotia and the Dairy Farmers of New Brunswick for generously providing access to their dairy farms, as well as for their valuable technical assistance and support throughout the multi-week data collection process. The authors also extend their gratitude to Jean Lynds, Operations Manager; Michael McConkey, Farm Manager; and Stewart Yuill, Animal Caretaker, for their unwavering support and assistance during data collection at the Ruminant Animal Centre, Dalhousie University.

## Data Availability Statement

The data is available upon reasonable request from the corresponding author.